\documentclass[usegraphicx,twocolumn,amsmath,amssymb]{mn2e} 
\def\nn{\nonumber} 
\usepackage{graphicx,amsmath,amssymb}

\title[Relativistic models of magnetars]{Relativistic models of
magnetars: the twisted-torus magnetic field configuration}

\author[R. Ciolfi {\it et al.}]
{R. Ciolfi$^1$, V. Ferrari$^1$, L. Gualtieri$^1$, J.A. Pons$^2$
\\
$^1$ Dipartimento di Fisica ``G.Marconi'',
Sapienza Universit\` a di Roma 
and Sezione INFN  ROMA1, 00185 Roma, Italy
\\ 
$^2$ Departament de F\'{\i}sica Aplicada, Universitat d'Alacant,
03080 Alacant, Spain }

\begin{document} 
\date{}

\maketitle

\begin{abstract} 
We find general relativistic solutions of equilibrium magnetic field
configurations in magnetars, extending previous results of Colaiuda
{\it et al.} \shortcite{colaiuda2008}.  Our method is based on the
solution of the relativistic Grad-Shafranov equation, to which
Maxwell's equations can be reduced.  We obtain equilibrium solutions
with the toroidal magnetic field component confined into a finite
region inside the star, and the poloidal component extending to the
exterior. These so-called twisted-torus configurations have been found
to be the final outcome of dynamical simulations in the framework of
Newtonian gravity, and appear to be more stable than other
configurations.  The solutions include higher order multipoles, which
are coupled to the dominant dipolar field. We use arguments of minimal
energy to constrain the ratio of the toroidal to the poloidal field.
\end{abstract} 


\section{Introduction}\label{introduction} 

In this paper we construct models of non rotating, strongly magnetized
neutron stars, or {\it magnetars} \cite{DT}, in general relativity.
We extend our previous work (Colaiuda {\it et al.}
\shortcite{colaiuda2008}) based on a formalism developed in Konno {\it
et al.}  \shortcite{KOK} and in Ioka \& Sasaki \shortcite{IS},
including the toroidal field in a {\it twisted-torus} configuration.
The extension to this field geometry is accomplished with an
appropriate choice of the function which determines, point by point,
the ratio between the toroidal and poloidal components of the magnetic
field.  The non-linear relation among the functions defining the
toroidal and poloidal fields, naturally leads to couplings between
different multipoles, thus making inadequate the one multipole
solution which is usually assumed.

A motivation to find consistent equilibrium solutions in general
relativity with this particular geometry comes from recent progresses
in numerical magneto-hydrodynamic (MHD) simulations, that have made
possible to study the dynamics and the stability of magnetic stars.
By following the time evolution of generic initial configurations, in
the framework of Newtonian gravity and using polytropic equations of
state, Braithwaite \& Spruit \shortcite{BS0} and Braithwaite \&
Nordlund \shortcite{BN} (see also Braithwaite \& Spruit
\shortcite{BS}) have found magnetic field configurations, which are
stable on timescales much longer than the Alfv\`en time: they decay
only due to finite resistivity.  These configurations are roughly
axisymmetric; the poloidal field extends throughout the entire star
and to the exterior, while the toroidal field is confined in a
torus-shaped region inside the star, where the field lines are closed.
These configurations were named {\it twisted-torus}.  Furthermore,
Yoshida {\it et al.}  \shortcite{yoshida2006} have shown that such
configurations are not significantly affected by rotation; Geppert \&
Rheinhardt \shortcite{GR06} studied the dependence of magnetostatic
equilibrium configurations on the rotational velocity and on the
initial angle between rotation and magnetic axis, finding hints for
the existence of a unique stable dipolar magnetostatic configuration,
independent of the initial field geometry.

We must remark that this particular field geometry resulting from
dynamical simulations is obtained assuming that outside the star there
is vacuum; consequently, outside the star electric currents are
forbidden and the magnetic field can only be poloidal.  This implies
that the toroidal field cannot extend to the exterior and that the
field lines which cross the surface are purely poloidal, whereas the
field lines confined inside the star can maintain a mixed (poloidal
and toroidal) structure.  This configurations appear to be stable on
dynamical timescales, probably due to magnetic helicity conservation,
which requires the persistence of a toroidal component of the field.
Notice, however, that different solutions including a magnetosphere
may be possible. In this case, the toroidal field could also extend to
the external region leading to a twisted magnetosphere
\cite{Lyu06,PTZN}.

In our perturbative approach, there is a free parameter which
represents the ratio between the toroidal and the poloidal components
of the magnetic field. We estimate the value of this parameter, by
identifying the configuration of minimal energy at fixed magnetic
helicity.  We also mention that in our configurations the external
field has mainly a dipole structure, with small corrections from
higher multipoles. Furthermore, confirming a previous suggestion by
Prendergast \shortcite{prendergast1956}, the toroidal and poloidal
fields have amplitudes of the same order of magnitude, whereas the
energy associated to the toroidal field is an order of magnitude
smaller than that of the poloidal field, since the former is confined
in a relatively small region.  Similar configurations have been found
in Newtonian models including rotation, in Yoshida \& Eriguchi
\shortcite{ye2006}, Yoshida {\it et al.}  \shortcite{yoshida2006}.

In this paper we consider non-rotating stars because observed
magnetars have a very slow rotation rate, although high rotation rates
may occur in the early stages of their evolution.

The structure of the paper is the following. The model is presented in
Section \ref{model}.  In Section \ref{dipole} we discuss a
configuration with purely dipolar magnetic field, neglecting the
couplings with higher multipoles.  In Section \ref{L1L2} we include
the $l=1$ and $l=2$ field components; Section \ref{multipoles}
accounts for the general case, including all multipoles and their
couplings. In Section \ref{helicity} we compute the total energy and
the magnetic helicity, and estimate the parameter $\zeta_0$ which
determines the ratio between toroidal and poloidal fields by energy
minimization; we also compute the magnetic energy, and compare the
contributions of the poloidal and toroidal fields for different values
of $\zeta_0$. In Section \ref{conclusions} we discuss the results and
draw the conclusions.
\section{Basic equations and formalism}\label{model}
We assume that the (non-rotating) magnetized star is stationary and
axisymmetric. We further assume that the magnetic field acts as a
perturbation of a spherically symmetric background describing a
spherical star.  The magnetized fluid is described within the
framework of ideal MHD, in which the effects of finite electrical
conductivity are neglected.  Rigorously speaking, this approximation
is only valid while the crust is still completely liquid and while the
core matter has not yet performed the phase transition to the
superfluid state, which is expected to occur at most a few hours after
birth (see e.g. section 5.1 in \cite{Aguilera08} and references
therein).  The onset of superfluidity and/or crystallization limits
the period during which magnetostatic equilibrium can be
established. Both the melting temperature and the critical
temperature of transition to the superfluid state, are between $10^9$
to $10^{10}$ K, and a typical neutron star quickly cools down below
this temperature in a few hours. However, since the characteristic
Alfv\'en time is on the order of $\tau_{{\rm A}}\approx 0.01-10$ s,
depending on the background field strength, there is ample time for
the magnetized fluid to reach a stable state, as shown in Braithwaite
\& Spruit (2006), while the state of matter is still liquid.  After
the crust is formed, the magnetic field is frozen in, and it only
evolves on a much longer timescale due to Ohmic dissipation or, in
some case, due to the Hall drift \cite{PG07}. Therefore, it is
reasonable to expect that the MHD equilibrium configurations set
within the first day after formation, will fix the magnetic field
geometry for a long time.

Here we first summarize the basic equations of ideal MHD in the
framework of general relativity and then introduce the perturbative
approach. Next, we obtain the form of the electromagnetic potential in
the case of twisted-torus configurations, and derive the relativistic
Grad-Shafranov equation.  We use spherical coordinates, $x^\mu
=(t,x^a,\phi)$, where $x^a =(r,\theta)$.  A stationary axisymmetric
space-time admits two killing vectors, $\eta=\partial/\partial t$ and
$\xi=\partial/\partial \phi$, and with our coordinate choice all
quantities (including the components of the metric tensor
$g_{\mu\nu}$) are independent of $t$ and $\phi$.

\subsection{Equations of ideal MHD in General Relativity}
According to a comoving observer with four-velocity $u^\mu$, the
stress-energy tensor of a perfect fluid with an electromagnetic field
is
\begin{equation}
T^{\mu\nu}=T^{\mu\nu}_{fluid}+T^{\mu\nu}_{em}
  \;\; , 
\label{new6}
\end{equation}
where 
\begin{equation}
T^{\mu\nu}_{fluid}=(\rho+P)u^{\mu}u^{\nu}+Pg^{\mu\nu} 
  \;\; , 
\label{new7}
\end{equation} 
\begin{equation}
T^{\mu\nu}_{em}=\frac{1}{4\pi}\left[\left( 
u^{\mu}u^{\nu}+\frac{1}{2}g^{\mu\nu}\right)B^2-
B^{\mu}B^{\nu}\right]  \;\; . 
\label{new8}
\end{equation} 
As usual, Euler's equations are found by projecting the equation
$T^{\mu\nu}_{\;\;\; ;\nu}=0$ orthogonally to $u^\mu$
\begin{equation} 
(\rho+P)a_{\mu}+P_{,\mu}+u_{\mu}u^{\nu}P_{,\nu}-f_{\mu}=0
  \;\; , 
\label{euler}
\end{equation}
where $f_{\mu}\equiv F_{\mu\nu} J^\nu$ is the Lorentz force and
$a_\mu=u^\nu u_{\mu;\nu}$ .  Here,
$F_{\mu\nu}=\partial_{\nu}A_{\mu}-\partial_{\mu}A_{\nu}$ is the
Maxwell tensor, in terms of which the electric and magnetic fields can
be defined as
\begin{equation}
E_{\mu}\equiv F_{\mu\nu}u^{\nu}\,,~~~~~
B_{\alpha}\equiv \frac{1}{2}\epsilon_{\alpha\beta\gamma\delta}\; 
u^{\beta} F^{\gamma\delta}\,.
\end{equation}

The basic equations of ideal, general relativistic MHD are, then: (i)
the continuity equation $(nu^{\mu})_{;\mu}=0$; (ii) Maxwell's
equations $F^{\mu\nu}_{\;\;\; ;\nu}=4\pi J^{\mu}$; (iii) the condition
of a vanishing electric field in the comoving frame
$E_{\mu}=F_{\mu\nu}u^{\nu}=0$ and (iv) Euler's equations
(\ref{euler}).

\subsection{The perturbative approach and the form of the 
electromagnetic potential}\label{pert}
We treat the magnetic field as an
axisymmetric perturbation of a spherically symmetric background 
and seek for stationary solutions.  The background metric is
\begin{equation} 
ds^2=-e^{\nu(r)}dt^2+e^{\lambda(r)}dr^2 +r^2 (d\theta^2+\sin^2{\theta}
d\phi^2) \;\; ,
\label{new10}
\end{equation}
where $\nu(r)$, $\mu(r)$ are solution of the unperturbed Einstein
equations (the TOV equations) for assigned equations of state. The
unperturbed 4-velocity is $u^\mu=(e^{-\nu/2},0,0,0)$.  To model the
unperturbed neutron star we use the equation of state of Akmal,
Pandharipande and Ravenhall called APR2 \cite{APR}, with a standard
equation of state for the stellar crust (see Benhar {\it et al.}
\shortcite{BFG}), which results in a neutron star of mass
$M=1.4\,M_{\odot}$ and radius $R=11.58$ km.  We remark that our EOS
accounts for the density-pressure relation in the crustal region, but
not for its elastic properties.  Our equations apply to a star where
the solid crust has not formed yet, or to configurations with a
relaxed crust where elasticity is irrelevant.

It can be shown (see for instance Colaiuda {\it et al.}
\shortcite{colaiuda2008}) that $(F_{\mu\nu},\,A_\mu,\,J^\mu)$ are of
order $O(B)$, and the perturbations $(\delta u^\mu,\,\delta
\rho,\,\delta P,\,\delta n,\,\delta g_{\mu\nu},\,\delta
G_{\mu\nu},\,\delta T_{\mu\nu})$ are of order $O(B^2)$.  Therefore, at
first order in $B$ the magnetic field is coupled only to the
unperturbed background metric (\ref{new10}), whereas the deformation
of the stellar structure induced by the magnetic field, which we do
not consider in this paper, appears at order $O(B^2)$.  Furthermore,
$(B^t,\,A_t,\,J^t,\,F_{t\nu})=O(B^3)$ and $(f_t,\,f_{\phi})=O(B^4)$.
Note that in the Grad-Shafranov equation, which we solve to order
$O(B)$, the metric perturbations do not appear; thus, to find the
magnetic field configurations we do not need to solve Einstein's
equations.  In Section \ref{helicity} and in Appendix \ref{energy}, we
will solve some components of Einstein's equations, in order to
evaluate the total energy of the system.

With these assumptions, the potential $A_\mu$, at $O(B)$, has the form
$A_\mu(r,\theta)=(0,A_r,A_\theta,A_{\phi})$. With an appropriate gauge
choice we can impose $A_\theta=0$ and write the potential as
\begin{equation} 
A_{\mu}=(0,e^{\frac{\lambda-\nu}{2}}\Sigma,0,\psi)
  \;\; , 
\label{new11}
\end{equation} 
where the components of $A_\mu$ are expressed in terms of two unknown
functions, $\Sigma(r,\theta)$ and $\psi(r,\theta)$.

A further simplification of $A_\mu$ is possible by exploiting the fact
that $f_\phi=-\psi_{,r}J^r-\psi_{,\theta}J^\theta=O(B^4)$. Using
Maxwell's equations and neglecting higher order terms, we find
\begin{equation} 
\tilde{\psi}_{,\theta}\psi_{,r}=\tilde{\psi}_{,r}\psi_{,\theta}
  \;\; , 
\label{new12}
\end{equation} 
where $\tilde{\psi}\equiv \sin{\theta}\,\Sigma_{,\theta}$. This result
implies $\tilde{\psi}=\tilde{\psi}(\psi)$ and allows us to write
\begin{equation}
\sin{\theta}\,\Sigma_{,\theta}=\zeta(\psi)\psi\label{sszp}\,,
\end{equation}
where $\zeta(\psi)$ is a function of $\psi$ of order $O(1)$.

The function $\zeta$ represents the ratio between the toroidal and
poloidal components of the magnetic field; different choices of this
function lead to qualitatively different field configurations.  The
simplest case is $\zeta=constant$ \cite{IS,colaiuda2008,Haskell}, but
with this choice (like with other simple choices of $\zeta(\psi)$) the
field lines of the toroidal field reach the exterior of the star,
where there is vacuum.  However, the magnetic field in vacuum can only
be poloidal (see, for instance, Colaiuda {\it et al.}
\shortcite{colaiuda2008}), thus this solution presents an
inconsistency.  To avoid this inconsistency, one should consider a
non-vacuum exterior, i.e. a magnetosphere, but modelling a neutron
star magnetosphere is a quite difficult task, especially in general
relativity. An alternative choice is to assume that the magnetic field
is entirely confined inside the star \cite{IS,Haskell}, but in this
way the parameter $\zeta$ must assume particular values; or,
one can instead accept that the toroidal field has a discontinuity at
the stellar surface, vanishing in the exterior \cite{colaiuda2008}; in
this way the entire range of $\zeta$ can be studied, but the
discontinuity in the field will, for consistence, imply the existence
of surface currents.

A different choice is made in this paper, where we assume that the
toroidal field is confined in a toroidal region inside the neutron
star, such that its field lines never cross the stellar surface, as in
the twisted-torus configuration. As mentioned in Section
\ref{introduction}, Newtonian numerical simulations \cite{BS0,BN,BS}
suggest that these configurations are indeed a quite generic outcome
of the evolution of strongly magnetized stars.

The twisted-torus configuration can be obtained by choosing the
following form of the function $\zeta$
\begin{equation}
\zeta(\psi)= \zeta_0 \left[ |\psi/\bar{\psi}|-1 \right]~
\Theta(|\psi/\bar{\psi}|-1)   \;\; .
\label{new13}
\end{equation}
A similar choice has been made, in a Newtonian framework, in Yoshida
{\it et al.} \shortcite{yoshida2006}.  In equation (\ref{new13}),
$\zeta_0$ is a constant of order $O(1)$; $\bar{\psi}$ is a constant of
order $O(B)$: it is the value of $\psi$ at the boundary of the
toroidal region where the toroidal field is confined (this boundary is
tangent to the stellar surface); finally,
$\Theta(|\psi/\bar{\psi}|-1)$ is the usual Heaviside function. With
this choice, the function $\zeta$ vanishes at the stellar surface,
where $r=R$, and the magnetic field
\begin{eqnarray}
B^\mu&=\Bigg( \;\; &0 \;\; , \;\;
\frac{e^{-\frac{\lambda}{2}}}{r^2\sin{\theta}}
\psi_{,\theta} \;\; , 
 \;\; -\frac{e^{-\frac{\lambda}{2}}}{r^2\sin{\theta}}\psi_{,r} \;\; , 
\nonumber\\
&&-\frac{e^{-\frac{\nu}{2}} \zeta_0 \psi\left( |\psi/\bar{\psi}|-1 
\right)}
{r^2\sin^2{\theta}}\,\Theta(|\psi/\bar{\psi}|-1) \;\; \Bigg)
\label{new14}
\end{eqnarray}
has no discontinuities.
\subsection{The relativistic Grad-Shafranov equation}\label{GSeq}
The Grad-Shafranov (GS) equation, which allows to determine the
magnetic field configuration, can be derived from the $\phi$-component
of Maxwell's equations
\begin{eqnarray}
J_{\phi}=-\frac{e^{-\lambda}}{4\pi}\left[\psi_{,rr}+\frac{\nu_{,r}
-\lambda_{,r}}{2}\psi_{,r} \right]
-\frac{1}{4\pi r^2}\left[\psi_{,\theta\theta}-\cot{\theta}
  \psi_{,\theta} 
\right]
\label{new14bis}
\end{eqnarray}
and from the $a$-components of Euler's equations (\ref{euler}), as
follows.  Euler's equations give
\begin{eqnarray} 
f_a&=&(\rho+P)a_a+P_{,a}+u_a u^{\nu}P_{,\nu}
\nonumber\\
&=&(\rho+P)\left(\frac{\nu}{2}-e^{\frac{\nu}{2}}
\delta u^t \right)_{,a}+P_{,a}+O(B^4)  \;\; . 
\label{new16bis}
\end{eqnarray}
For barotropic equations of state $P=P(\rho)$, the first principle of
thermodynamics allows to write
\begin{equation}
P_{,a}=(\rho+P)\left(\ln\frac{\rho+P}{n}\right)_{,a}
  \;\; ,
\label{new17}
\end{equation}
then (\ref{new16bis}) yields
\begin{equation} 
f_a=(\rho+P)\chi_{,a}
  \;\; , 
\label{new17bis}
\end{equation}
where $\chi=\chi(r,\theta)$. On the other hand, the $a$-components of
the Lorentz force $f_\mu=F_{\mu\nu}J^\nu$ can be written as (see
Colaiuda {\it et al.} \shortcite{colaiuda2008})
\begin{equation}
f_a=\frac{\psi_{,a}}{r^2 \sin^2{\theta}}\tilde{J}_{\phi}
  \;\; ,
\label{new15}
\end{equation}
where, in the present case,  
\begin{equation}
\tilde{J}_{\phi}=J_\phi-\frac{e^{-\nu}\zeta_0^2}{4\pi}
[\psi-3\psi |\psi/\bar{\psi}|+2\psi^3/\bar{\psi}^2]
~\Theta(|\psi/\bar{\psi}|-1)
  \;\; .
\nonumber
\end{equation}
Therefore, 
\begin{equation}
\frac{\psi_{,a}}{r^2 \sin^2{\theta}}\tilde{J}_{\phi}=(\rho+P)\chi_{,a}\,.
\end{equation}
From $\chi_{,r\theta}-\chi_{,\theta r}=0$ it follows that 
\begin{equation}
\psi_{,r}\left(\frac{\tilde{J}_{\phi}}{(\rho+P)r^2
  \sin^2{\theta}}\right)_{,\theta}
-\psi_{,\theta}\left(\frac{\tilde{J}_{\phi}}{(\rho+P)r^2
  \sin^2{\theta}}\right)_{,r} =0 \;\; ,
\nonumber
\end{equation}
which implies 
\begin{equation}
\left(\frac{\tilde{J}_{\phi}}{(\rho+P)r^2
  \sin^2{\theta}}\right)=F(\psi)=c_0+c_1\psi+O(B^2) \;\; ,
\label{new18bis}
\end{equation}
with $c_0$, $c_1$ constants of order $O(B)$, $O(1)$ respectively.
Hence, $J_{\phi}$ turns out to be
\begin{eqnarray}
J_\phi&=&\frac{e^{-\nu}\zeta_0^2}{4\pi}[\psi-3\psi
|\psi/\bar{\psi}|+2\psi^3/\bar{\psi}^2]
~\Theta(|\psi/\bar{\psi}|-1)
\nonumber\\
&&+(\rho+P)r^2 \sin^2{\theta}\,
[c_0+c_1\psi] \;\; .
\label{new19}
\end{eqnarray}
From Eqns. (\ref{new14bis}), (\ref{new19}) the relativistic GS
equation at first order in $B$ is finally obtained:
\begin{eqnarray}
&&-\frac{e^{-\lambda}}{4\pi}\left[\psi_{,rr}+\frac{\nu_{,r}
-\lambda_{,r}}{2}\psi_{,r}\right] -\frac{1}{4\pi r^2}
\left[\psi_{,\theta\theta}-\cot{\theta}
\psi_{,\theta} \right] 
\nonumber\\
&&-\frac{e^{-\nu}\zeta_0^2}{4\pi}
\left[\psi-3\psi |\psi/\bar{\psi}|+2\psi^3/\bar{\psi}^2 \right]
~\Theta(|\psi/\bar{\psi}|-1)
\nonumber\\ 
&&=(\rho+P)r^2\sin^2{\theta}\, [c_0+c_1\psi] \;\; .
\label{new20}
\end{eqnarray}
If we now define
$\psi(r,\theta)\equiv \sin{\theta}a(r,\theta)_{,\theta}$ and expand
the function $a(r,\theta)$ in Legendre polynomials
\begin{equation}
a(r,\theta)=\sum_{l=1}^{\infty} a_l(r)P_l(\cos{\theta})\;\; ;
\label{new21}
\end{equation}
the GS equation rewrites as
\begin{eqnarray}
&&-\frac{\sin{\theta}}{4\pi}\sum_{l=1}^{\infty}P_{l,\theta}
\left(e^{-\lambda}a_l''
+e^{-\lambda}\frac{\nu'-\lambda'}{2}a_l'-\frac{l(l+1)}{r^2}a_l \right)
\nonumber\\ 
&&-\frac{e^{-\nu}}{4\pi}
\;\Theta\left(\left|\frac{1}{\bar{\psi}}\sum_{l=1}^{\infty} a_l
P_{l,\theta}\sin{\theta}\right|-1 \right)\; \zeta_0^2\Bigg[
\sum_{l=1}^{\infty} a_l
P_{l,\theta}\sin{\theta}
\nonumber\\
&&-\frac{3}{|\bar{\psi}|}\left(
\sum_{l,l'=1}^{\infty} a_l P_{l,\theta}\sin{\theta} |a_{l'}
P_{l',\theta}\sin{\theta}| \right) \nonumber\\
&&+\frac{2}{\bar{\psi}^2} \left( \sum_{l,l',l''=1}^{\infty} a_l a_{l'}
a_{l''} P_{l,\theta} P_{l',\theta} P_{l'',\theta} \sin^3{\theta}
\right) \Bigg] 
\nonumber\\
&&=(\rho+P)r^2 \sin^2{\theta} \left[c_0+c_1
\sum_{l=1}^{\infty} a_l P_{l,\theta}\sin{\theta}\right] \;\; .
\label{new22}
\end{eqnarray}
Here and in the following we denote with primes the differentiation
with respect to $r$.

Finally, projecting Eq. (\ref{new22}) onto the different harmonic
components, we obtain a system of coupled ordinary differential
equations for the functions $a_l(r)$. The projection is performed
using the property
\begin{equation}
\frac{2l'+1}{2l'(l'+1)} \int_0^\pi P_{l,\theta} P_{l',\theta} 
\sin{\theta} \;d\theta\;
=\delta_{ll'}  \;\; .
\label{new23}
\end{equation}
If we consider the contribution of $n$ different harmonics, we need to
solve a system of $n$ coupled equations, obtained from (\ref{new22}),
for the $n$ functions $a_l(r)$.

\subsection{Boundary conditions}\label{boundL1}
The functions $a_l(r)$ must have a regular behaviour at the origin; by
taking the limit $r\rightarrow 0$ of the GS equation one can find
\begin{equation}
a_l(r\rightarrow 0)=\alpha_l r^{l+1}\;,
\label{alphal}
\end{equation}
where $\alpha_l$ are arbitrary constants to be fixed. 

Outside the star, where there is vacuum and the field is purely poloidal,
equations (\ref{new22}) decouple, and  can be solved 
analytically  for each value of $l$. The solution
can be expressed in terms of the generalized hypergeometric functions
($F([l_1,l_2],[l_3],z)$), also known as Barnes' extended
hypergeometric functions, as follows:
\begin{eqnarray}
a_l&=&A_1 ~r^{-l} ~F([l,l+2], [2+2l], z)
\nonumber\\
&&+A_2 ~r^{l+1} ~ F([1-l,-1-l], [-2l], z)\;~.
\label{outerPhi}
\end{eqnarray}
where $z=2 M/r$ and $A_1$ and $A_2$ are arbitrary integration
constants, which must be fixed according to the values of the magnetic
multipole moments. Regularity of the external solution at $r=\infty$
requires $A_2=0$ for all multipoles. 
For example, for $l=1,2,3$ we have
\begin{eqnarray}
a_1 &\propto& r^2 \left[ \ln(1-z)
+ z + \frac{z^2}{2}\right] 
\nonumber \\
a_2 &\propto& r^3 \bigg[ (4-3z) \ln(1-z)
+ 4z - {z^2} - \frac{z^3}{6}\bigg] 
\nonumber \\
a_3 &\propto& r^4 \bigg[ (15-20z+6 z^2) \ln(1-z)+ 15 z
\nonumber\\ 
&&\qquad- \frac{25 z^2}{2} + z^3 + \frac{z^4}{12}\bigg]\;\;.
\label{vacsol}
\end{eqnarray}
At the stellar surface  we require the field to be continuous.
This condition is satisfied if $a_l$ and $a_l'$ are continuous.
For practical purposes, the boundary
conditions at $r=R$ can be written as
\begin{equation}
a_l' = -\frac{l}{R} f_l a_l
\label{OBC}
\end{equation}
where $f_l$ is a relativistic factor which only depends on the star
compactness  $2M/R$ (in the Newtonian limit all $f_l=1$), and can
be numerically evaluated  with the help of any algebraic
manipulator. For our model ($2M/R=0.357$), the values of $f_l$ for the
first five multipoles are 1.338, 1.339, 1.315, 1.301, and 1.292,
respectively.

In general, there are $n+2$ arbitrary constants to be fixed: the $n$
constants $\alpha_l$, associated to the condition (\ref{alphal}),
$c_0$ and $c_1$.  Thus, we need to impose $n+2$ constraints, of which
$n+1$ are determined by the boundary conditions.  $n$ conditions are
provided by Eq. (\ref{OBC}), i.e. by imposing continuity in $r=R$ of
the ratios $a_l'/a_l$. The overall normalization of the field gives
another condition, which is fixed by imposing that the value of the
$l=1$ contribution at the pole is $B_{pole}=10^{15}$ G (this
corresponds to set $a_1(R)=1.93\cdot 10^{-3}$ km). The reason for this
choice is that the surface value of the magnetic field is usually
inferred from observations by applying the spin-down formula, and
assuming a purely dipolar external field; for magnetars, the value of
$B$ estimated in this way is $\sim10^{14}-10^{15}$ G.  The remaining
condition will be imposed as follows.

In the case  of a purely dipolar field ($l=1$),
we shall assume $c_1=0$.
In the general multipolar case, 
we choose to impose that the external contribution
of all the $l>1$ harmonics, i.e. $\sum_{l>1}a_{l}(R)^2$,
is minimum.  

\begin{figure*}
\centering
\begin{minipage}{176mm}
\begin{center}
\includegraphics[width=4cm,angle=270]{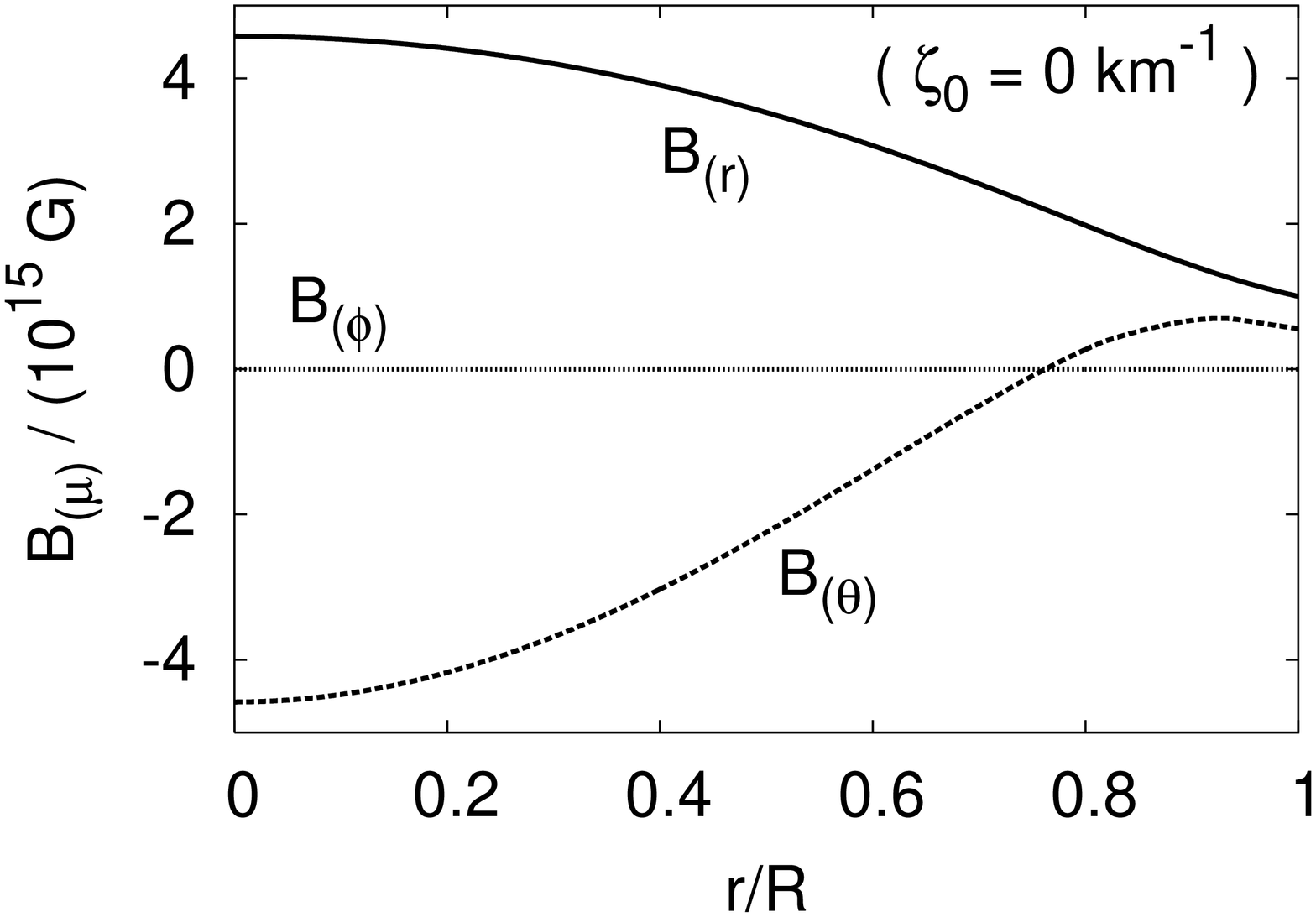}
\includegraphics[width=4cm,angle=270]{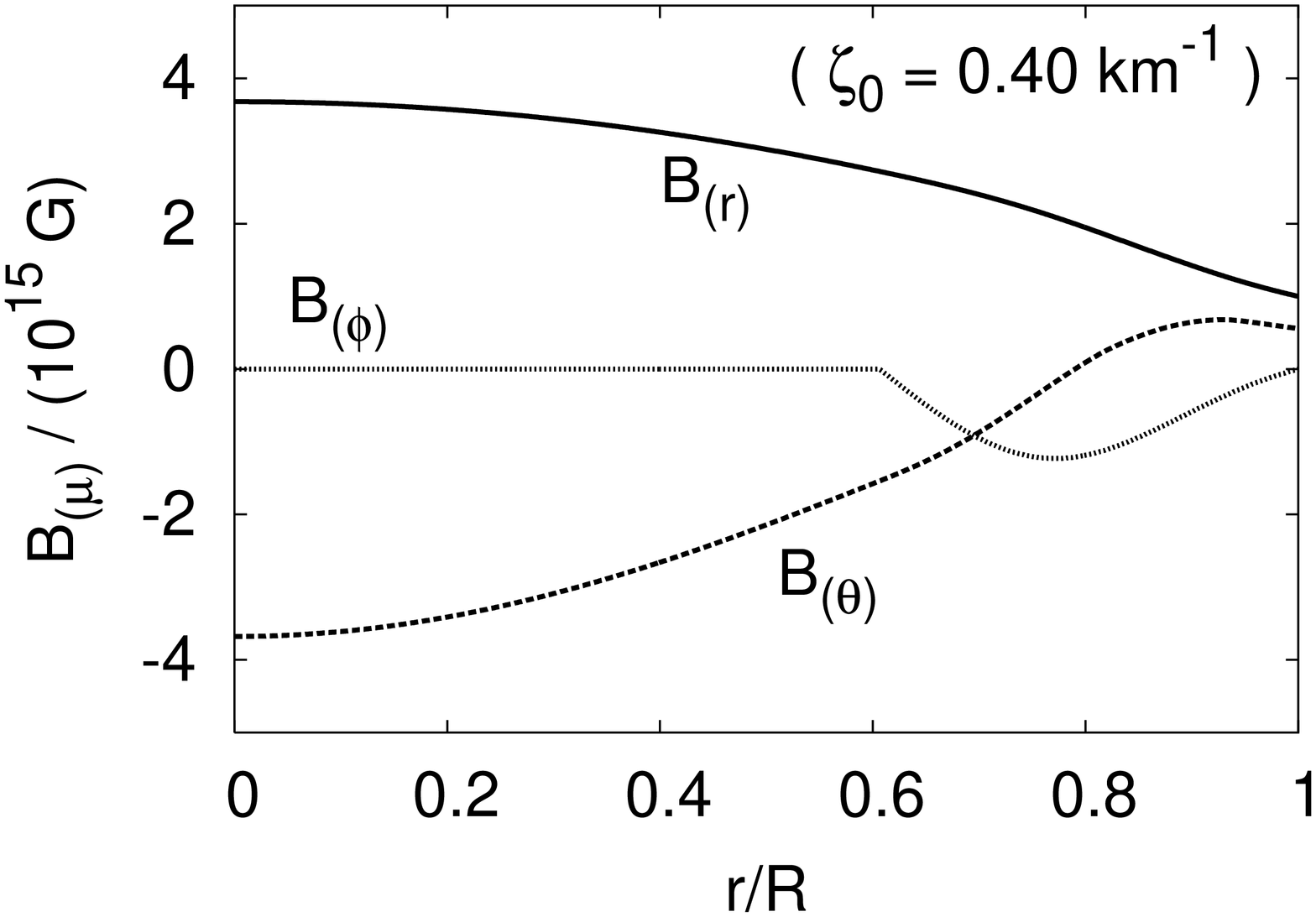}
\includegraphics[width=4cm,angle=270]{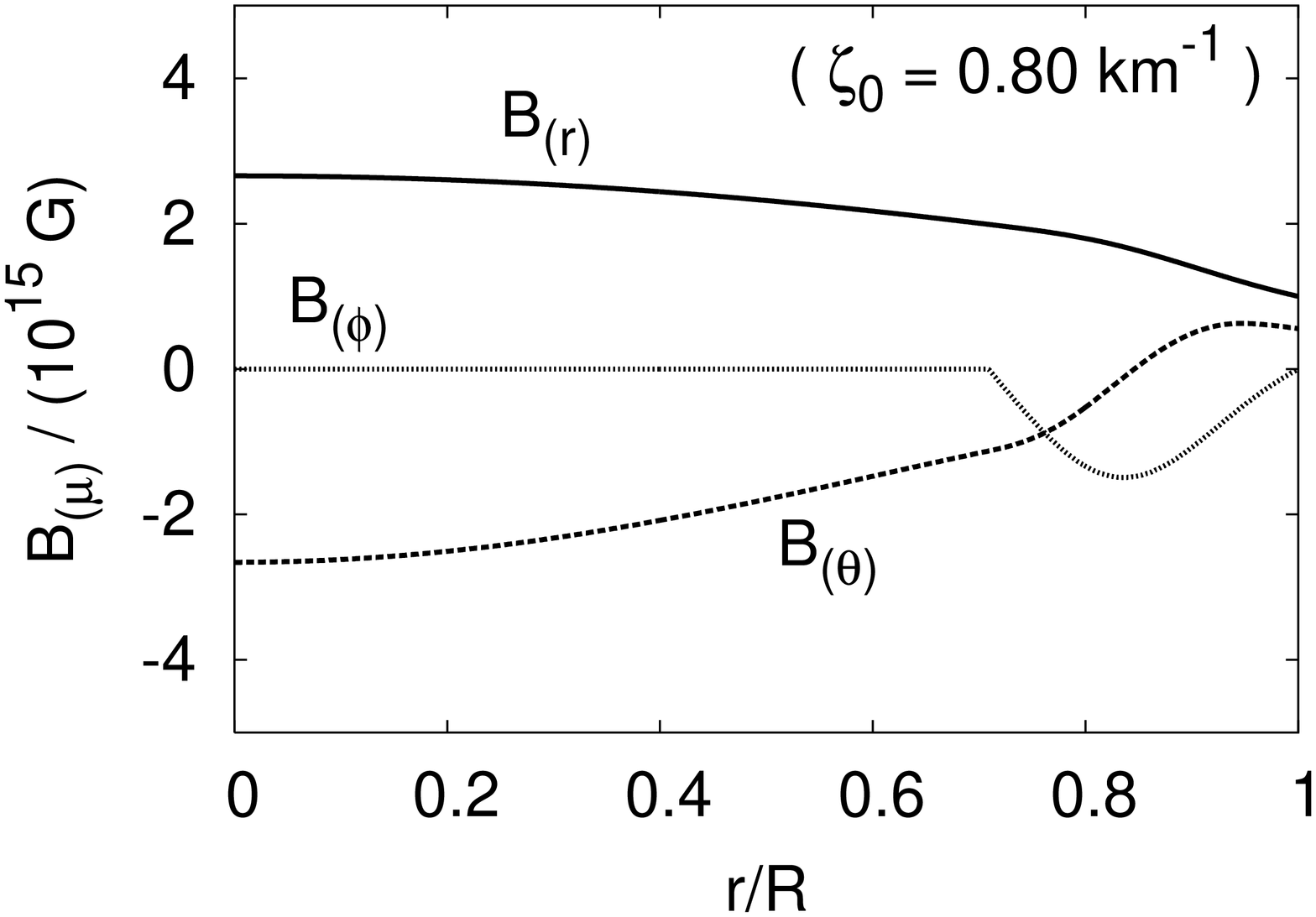}
\end{center}
\caption{The profiles of the tetrad components of the magnetic field
($B_{(r)}(\theta=0)$, $B_{(\theta)}(\theta=\pi/2)$,
$B_{(\phi)}(\theta=\pi/2)$) are shown for the purely dipolar case
with $\zeta_0=0$ km$^{-1}$, $\zeta_0=0.40$ km$^{-1}$ and
$\zeta_0=0.80$ km$^{-1}$.
\label{fig1}}
\end{minipage}
\end{figure*}
\section{The case of purely dipolar field}\label{dipole}
We begin discussing the simplest case of a purely dipolar
configuration, in which all couplings with higher order multipoles are
neglected in Eq. (\ref{new22}) ($a_{l>1}=0$).  In this case, for any
assigned value of $\zeta_0$ there exists an infinite set of solutions,
each corresponding to a value of $c_1$; these solutions describe
qualitatively similar magnetic field configurations.

However, when higher order harmonics are taken into account, as we
will see in the next Section, the picture changes.  For instance, when
$\zeta_0=0$ and the $l=1,2$ harmonic components are included, the
equations for $a_1$ and $a_2$ decouple: the equation for $a_1$ is the
same as in the purely dipolar case, but a solution for $a_2$
satisfying the appropriate boundary conditions exists only for a
unique value of $c_1$.  Therefore, in the general case $c_1$ is not a
truly free parameter (this is true also for $\zeta_0\neq 0$), and the
fact that in the purely dipolar case it looks as such, is an artifact
of the truncation of the $l>1$ multipoles.  In order to provide a
mathematically simple example, which will be useful to understand the
structure of the twisted-torus configurations, in this Section we
shall consider the simplest case $c_1=0$.

By projecting Eq. (\ref{new22}) onto the $l=1$ harmonic, 
and neglecting all contributions from $l>1$ terms
we find
\begin{eqnarray}
&&\frac{1}{4\pi}\left(e^{-\lambda}a_1''
+e^{-\lambda}\frac{\nu'-\lambda'}{2}a_1'-\frac{2}{r^2}a_1 \right)
\nonumber\\
&&-\frac{e^{-\nu}}{4\pi} \int_0^\pi (3/4) 
\;\Theta\left(\left| \frac{-a_1\sin^2{\theta}}{\bar{\psi}}\right|-1 \right)\; 
\nonumber\\
&&\cdot\,\zeta_0^2\Bigg[-a_1+3 a_1\left|
\frac{-a_1\sin^2{\theta}}{\bar{\psi}}\right|-2a_1^3
\sin^4{\theta}
/\bar{\psi}^2 \Bigg] \sin^3{\theta}\;d\theta\;
\nonumber\\
&&=(3/4)\int_0^\pi c_0 (\rho+P)r^2 \sin^3{\theta}\;d\theta\;
=c_0 (\rho+P)r^2   \;\; .
\label{new24}
\end{eqnarray}
The tetrad components of the magnetic field (i.e. the components
measured by a locally inertial observer) are:
\begin{eqnarray}
B_{(r)}&=&\frac{\psi_{,\theta}}{r^2\sin{\theta}} \;\; ,
\nonumber\\
B_{(\theta)}&=&-\frac{e^{-\frac{\lambda}{2}}}{r\sin{\theta}}\psi_{,r}
  \;\; ,\nonumber\\
B_{(\phi)}&=&-\frac{e^{-\frac{\nu}{2}} 
\zeta_0 \psi\left( |\psi/\bar{\psi}|-1 \right)}
{r \sin{\theta}}\cdot\Theta(|\psi/\bar{\psi}|-1)  \;\; ,
\label{new24a}
\end{eqnarray} 
where $\psi=-a_1\sin^2{\theta}$.

\begin{figure}
\hspace{1.8cm}
\includegraphics[width=8cm,angle=0]{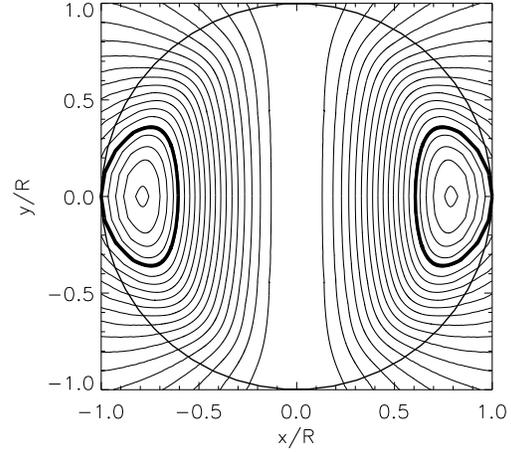}
\vspace{0.2cm}
\caption{The projection of the field lines in the meridional plane are
shown for the purely dipolar case with $\zeta_0=0.40$ km$^{-1}$. The
toroidal field is confined within the marked region.
\label{fig2}}
\end{figure}

The profiles of the tetrad components of the field inside the star,
are plotted in Fig. \ref{fig1} for increasing values of $\zeta_0$;
$B_{(r)}$ is evaluated in $(\theta=0)$ and $B_{(\theta)}$,
$B_{(\phi)}$ in $(\theta=\pi/2)$.  In Fig. \ref{fig2} we show the
projection of the field lines in the meridional plane, for
$\zeta_0=0.40$ km$^{-1}$.  Figs. \ref{fig1} and \ref{fig2} show that
the toroidal field $B_{(\phi)}$ is confined within a torus-shaped
region; its amplitude ranges from zero, at the border of the region,
to a maximum, close to its center.  At the stellar surface and in the
exterior $B^\phi$ vanishes, showing that there is no discontinuity in
the toroidal field. The panels of Fig. \ref{fig1} show the field
profiles for different values of $\zeta_0$: larger values of $\zeta_0$
correspond to a toroidal field with increasing amplitude, confined in
an increasingly narrow region close to the stellar surface, while the
amplitude of the poloidal components ($B_{(r)}$, $B_{(\theta)}$)
decreases. We remark that this implies that inside the star we cannot
have a twisted-torus configuration where the toroidal component
dominates with respect to the poloidal one: if $|B_{(\phi)}|$ becomes
larger with respect to $|B_{(r)}|$ and $|B_{(\theta)}|$, the domain
where it is non vanishing shrinks.

\section{The case with \MakeLowercase{$l=1$} and \MakeLowercase{$l=2$} 
multipoles}\label{L1L2}
\begin{figure*}
\centering
\begin{minipage}{176mm}
\begin{center}
\hskip 1.7cm
\includegraphics[width=5.9cm,angle=0]{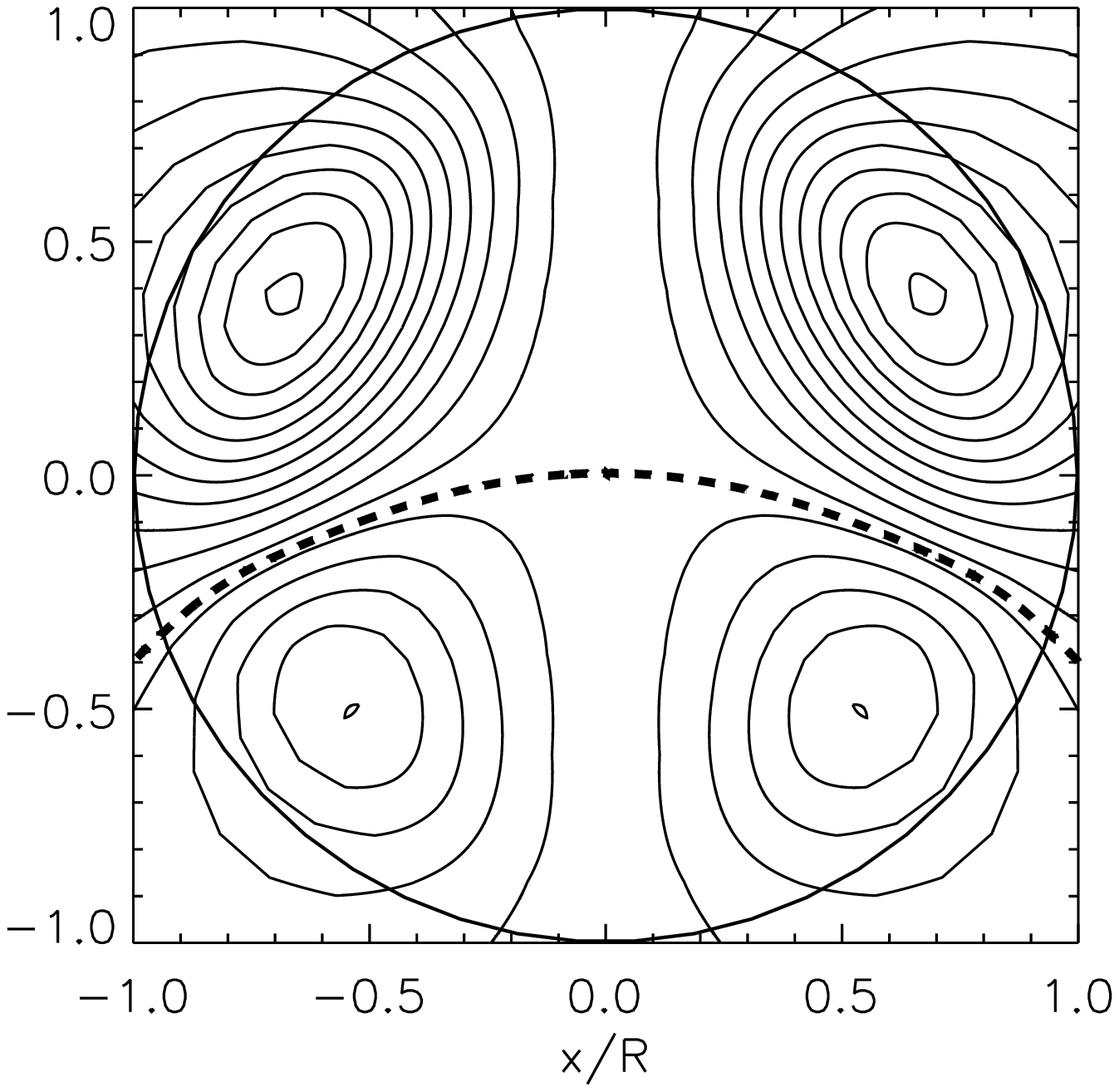}
\hskip -1cm
\includegraphics[width=5.9cm,angle=0]{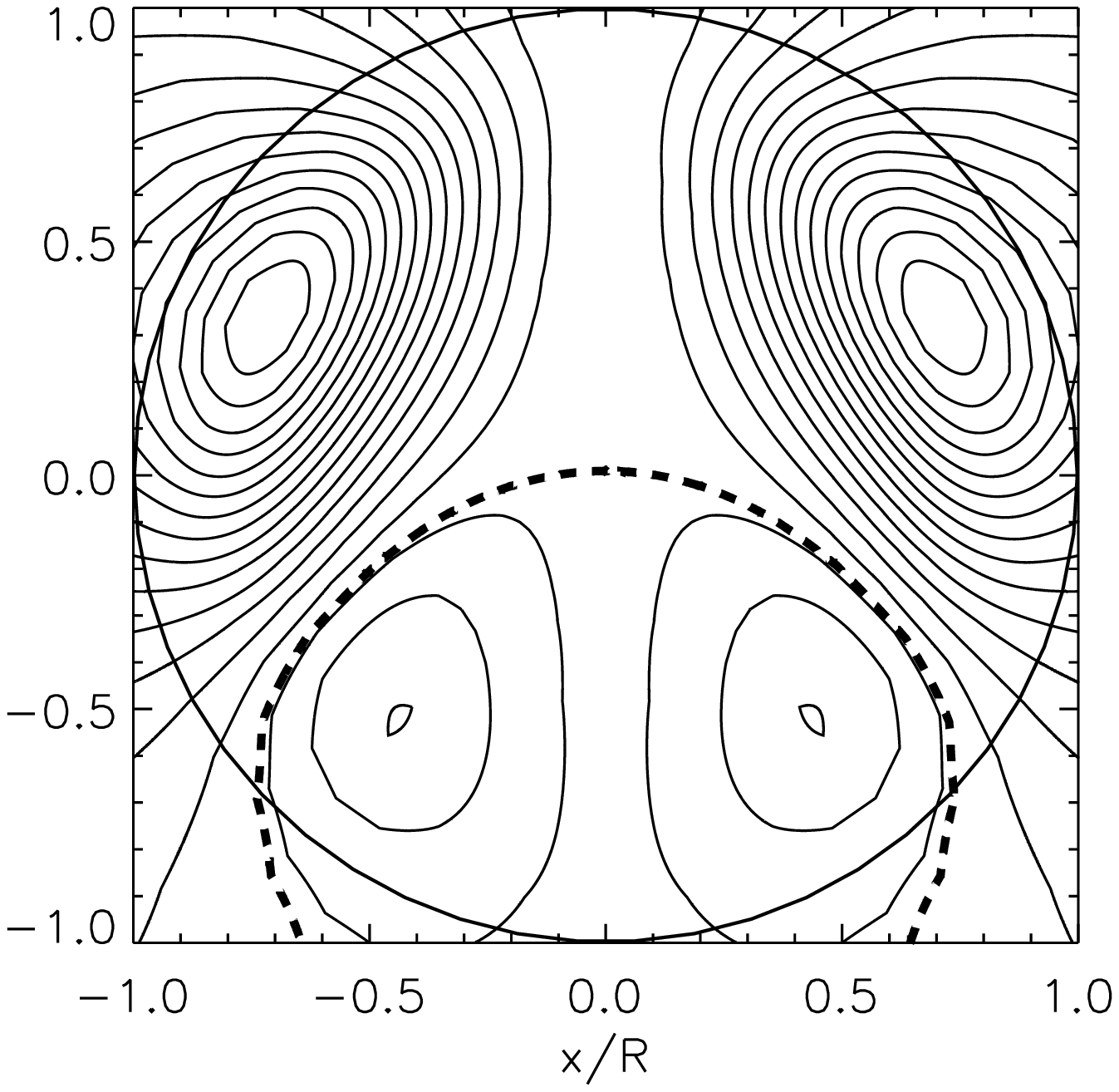}
\hskip -1cm
\includegraphics[width=5.9cm,angle=0]{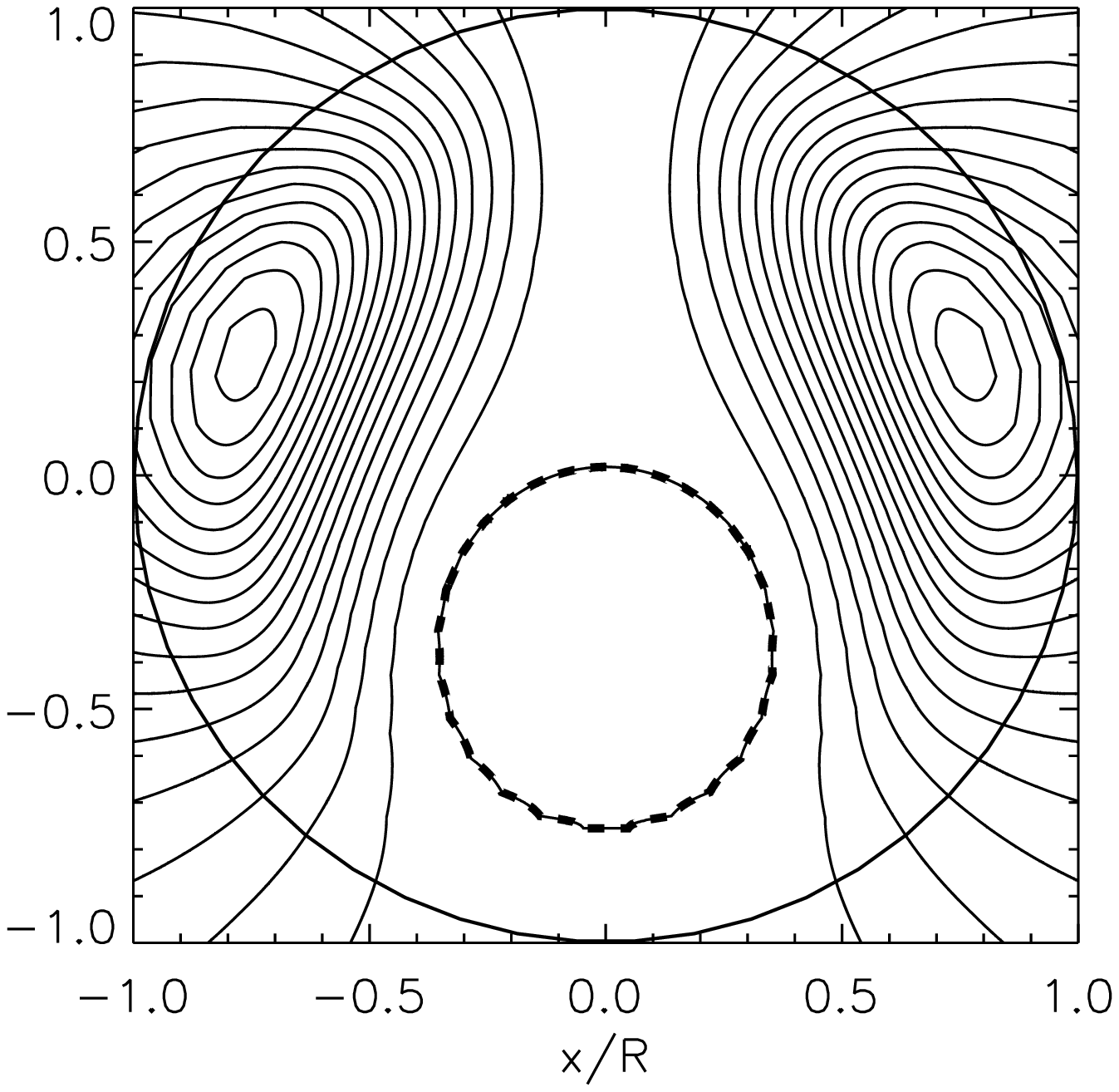}
\end{center}
\caption{The projection of the field lines in the meridional plane are
shown for $\zeta_0=0$ km$^{-1}$ and $a_2(R)/a_1(R)=1,1/2,1/4$
respectively, and for $a_{l>2}=0$.  The dashed line corresponds to
$\psi=0$.
\label{fig3}}
\end{minipage}
\end{figure*}
We now proceed with our investigation considering the $l=1$ and $l=2$
contributions, and setting $a_{l>2}=0$. The projection of the GS equation
(\ref{new22}) onto the harmonics $l=1$ and $l=2$ gives the following
coupled equations

\begin{eqnarray}
&&\frac{1}{4\pi}\left(e^{-\lambda}a_1''
+e^{-\lambda}\frac{\nu'-\lambda'}{2}a_1'-\frac{2}{r^2}a_1 \right)
\nonumber\\
&&-\frac{e^{-\nu}}{4\pi} \int_0^\pi (3/4) 
\;\Theta\left( \left|\frac{-a_1-3 a_2 
\cos{\theta}}{\bar{\psi}}\sin^2{\theta}\right|-1 \right)
\nonumber\\
&&\cdot\;\zeta_0^2 \Bigg[-a_1-3 a_2 \cos{\theta}
+3(a_1+3 a_2 \cos{\theta})\nonumber\\
&&\cdot\left|\frac{-a_1-3 a_2 
\cos{\theta}}{\bar{\psi}}\sin^2{\theta}\right|
+2\sin^4{\theta}\big(-a_1^3-9 a_1^2 a_2 \cos{\theta}\nonumber\\
&&-27a_1 a_2^2\cos^2{\theta}
-27 a_2^3 \cos^3{\theta}\big)/\bar{\psi}^2
\Bigg] \sin^3{\theta}\;d\theta \nonumber\\
&&=(\rho+P)r^2 \left(c_0-\frac{4}{5}c_1 a_1\right) \;\; ,
\label{new27}
\end{eqnarray}
\begin{eqnarray}
&&\frac{1}{4\pi}\left(e^{-\lambda}a_2''
+e^{-\lambda}\frac{\nu'-\lambda'}{2}a_2'-\frac{6}{r^2}a_2 \right)\nn\\
&&+\frac{e^{-\nu}}{4\pi} \int_0^\pi (5/12) \;
\Theta\left( \left|\frac{-a_1-3 a_2 
\cos{\theta}}{\bar{\psi}}\sin^2{\theta}\right|-1 \right)\;\nonumber\\
&&\cdot\zeta_0^2 \Bigg[-a_1-3 a_2 \cos{\theta}
+3(a_1+3 a_2\cos{\theta})\nn\\
&&\cdot\left|\frac{-a_1-3 a_2
\cos{\theta}}{\bar{\psi}}\sin^2{\theta}\right|
+2\sin^4{\theta}\big( -a_1^3-9 a_1^2 a_2 \cos{\theta}\nonumber\\
&&-27a_1 a_2^2
\cos^2{\theta}-27 a_2^3 \cos^3{\theta}\big)/\bar{\psi}^2 \Bigg]
(-3\cos{\theta}\sin^3{\theta})\;d\theta\; \nonumber\\
&&=-\frac{4}{7}(\rho+P)r^2 c_1 a_2  \;\; .\label{new28}
\end{eqnarray}
We integrate this system by imposing the boundary conditions discussed
above, i.e. a regular behaviour at the origin (equation
(\ref{alphal})) and continuity at the surface of $a_1,a_1',a_2,a_2'$
with the analytical external solutions given by (\ref{vacsol}).

Let us first consider the simple case $\zeta_0=0$.
Eqns. (\ref{new27}), (\ref{new28}) decouple, and become
\begin{eqnarray}
&&e^{-\lambda}a_1''+e^{-\lambda}\frac{\nu'-\lambda'}{2}
a_1'-\frac{2}{r^2}a_1\nonumber\\
&&=4\pi (\rho+p)r^2\left[c_0-\frac{4}{5}c_1a_1\right]\nonumber\\
&&e^{-\lambda}a_2''+e^{-\lambda}\frac{\nu'-\lambda'}{2}
a_2'-\frac{6}{r^2}a_2\nonumber\\
&&=-\frac{16\pi}{7}(\rho+p)r^2c_1a_2\,.\label{sysdec}
\end{eqnarray}
There are four constants to fix ($\alpha_1,\alpha_2,c_0,c_1$) and
three conditions: $a_1(R)=1.93\cdot 10^{-3}$ km (normalization) and
the ratios $a_1'(R)/a_1(R)$ and $a_2'(R)/a_2(R)$ from the matching
with the exterior solutions; thus, we need an additional requirement.
We remark that we cannot impose $c_1=0$ as in the purely dipolar case,
because the ratio $a_2'(R)/a_2(R)$ depends only on $c_1$, and the
matching with the exterior solution is possible only for a particular
value of $c_1$, i.e. $c_1=0.84$ km$^{-2}$.

If we impose that $|a_2(R)|$ is minimum, we find that this condition
yields the trivial solution $a_2(r)\equiv0$ (with non-vanishing
$a_1$). Indeed, from Eqns. (\ref{sysdec}) it is straightforward to see
that $a_2(r)\equiv0$ is a solution of the system.  When $\zeta_0\neq
0$, equations (\ref{new27}), (\ref{new28}) are coupled, but they still
allow the trivial solution $a_2(r)\equiv0$, which minimizes
$|a_2(R)|$, with non-vanishing $a_1$.  The existence of this solution
is a remarkable property of this system, and it is due to the fact
that the integral in $\theta$ on the left-hand side of
Eq. (\ref{new28}) vanishes for $a_2=0$ (the integrand becomes odd for
parity transformations $\theta\rightarrow\pi-\theta$).  Hence, if we
look for a solution which minimizes the contributions from the $l>1$
components at the stellar surface, we have to choose the trivial
solution $a_2(r)\equiv 0$.

If, instead, we do not require that $a_2(R)$ is minimum, and we assign
a finite value to the ratio $a_2(R)/a_1(R)$, we find a non-trivial
field configuration which is non symmetric with respect to the
equatorial plane.  This feature is shown in Fig. \ref{fig3}, where the
projection of the field lines in the meridional plane are plotted for
$\zeta_0=0$ and $a_2(R)/a_1(R)$ equal to $1$, $1/2$ and $1/4$,
respectively.

\section{The general case}\label{multipoles}
When all harmonics are taken into account, there exist two distinct
classes of solutions: those {\it symmetric} (with respect to the
equatorial plane), with vanishing even order components
($a_{2l}\equiv0$), and the {\it antisymmetric} solutions, with
vanishing odd order components ($a_{2l+1}\equiv0$).  Both solutions
satisfy the GS equation (\ref{new22}).  Let us consider the symmetric
class. When $a_{2l}=0$ the integrals arising when equation
(\ref{new22}) is projected onto the even harmonics, which couple odd
and even terms, vanish since the integrands change sign under parity
transformations.  Therefore, the symmetric solutions can be found by
setting $a_{2l}\equiv0$, projecting Eq. (\ref{new22}) onto the odd
harmonics and solving the resulting equations for $a_{2l+1}$.
Similarly, the integrals in equation (\ref{new22}) projected onto the
odd harmonics, vanish when $a_{2l+1}=0$; thus, we can consistently set
$a_{2l+1}\equiv0$, and find the antisymmetric solutions using the same
procedure.
\begin{figure*}
\centering
\begin{minipage}{176mm}
\begin{center}
\includegraphics[width=4cm,angle=270]{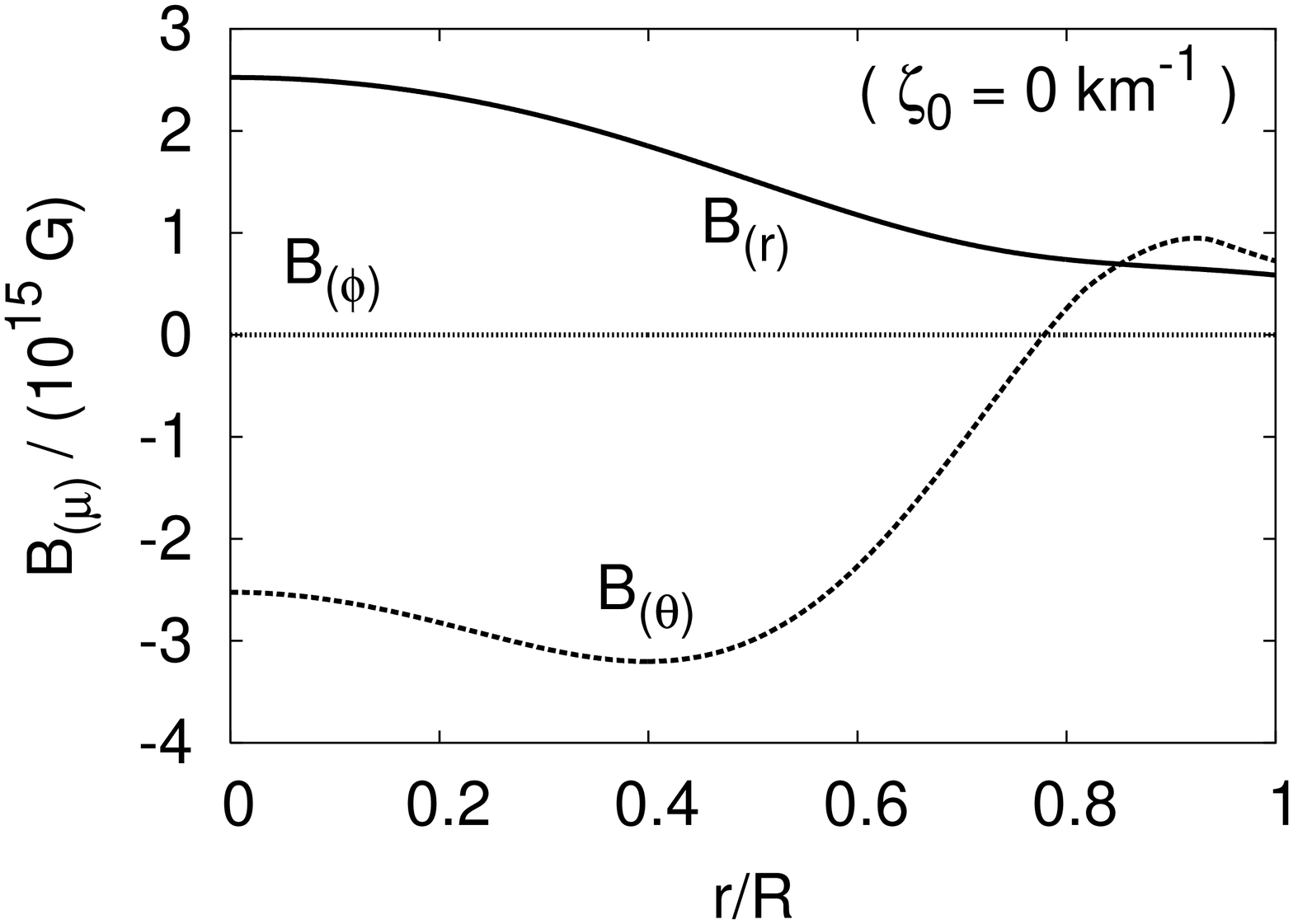}
\includegraphics[width=4cm,angle=270]{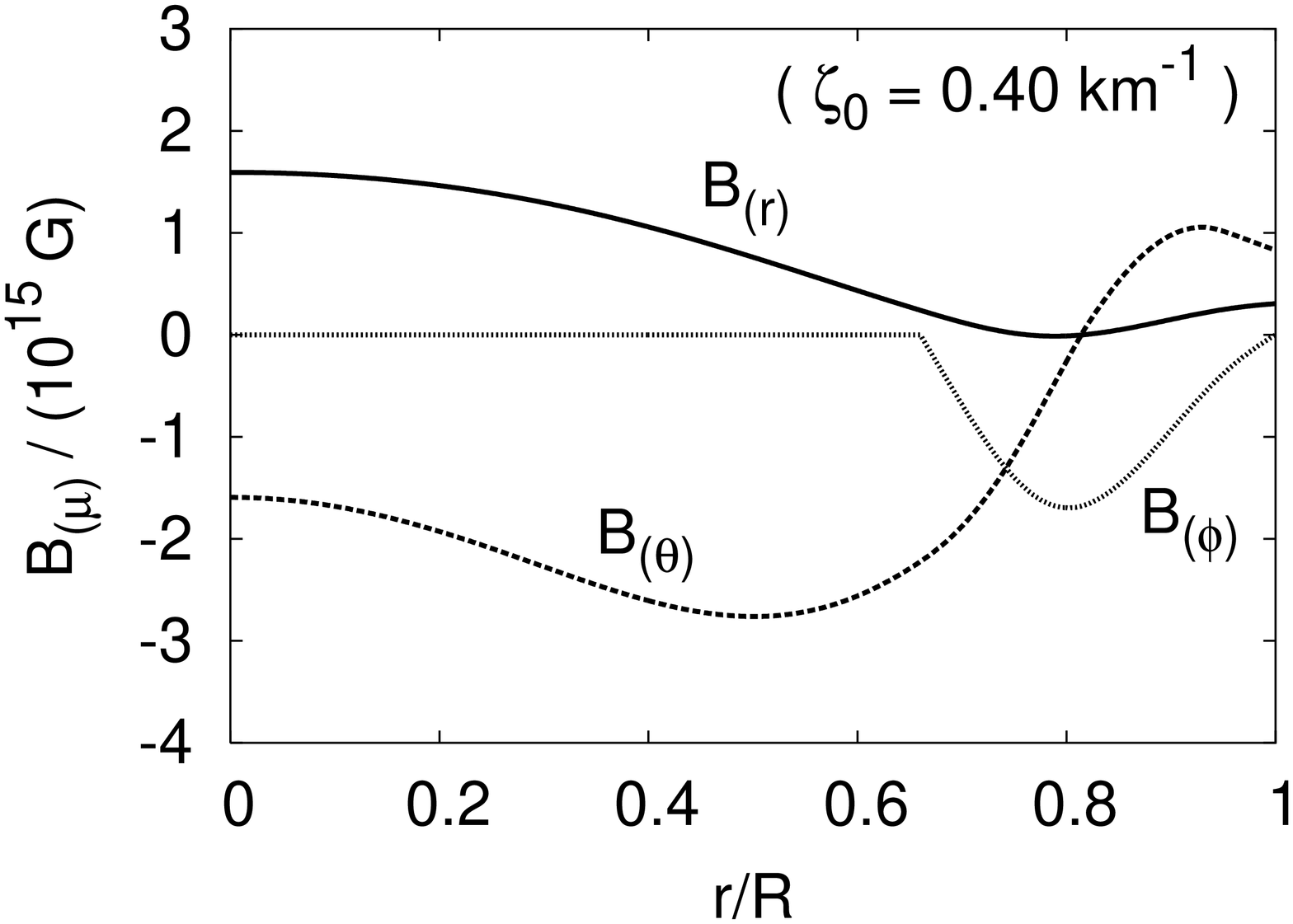}
\includegraphics[width=4cm,angle=270]{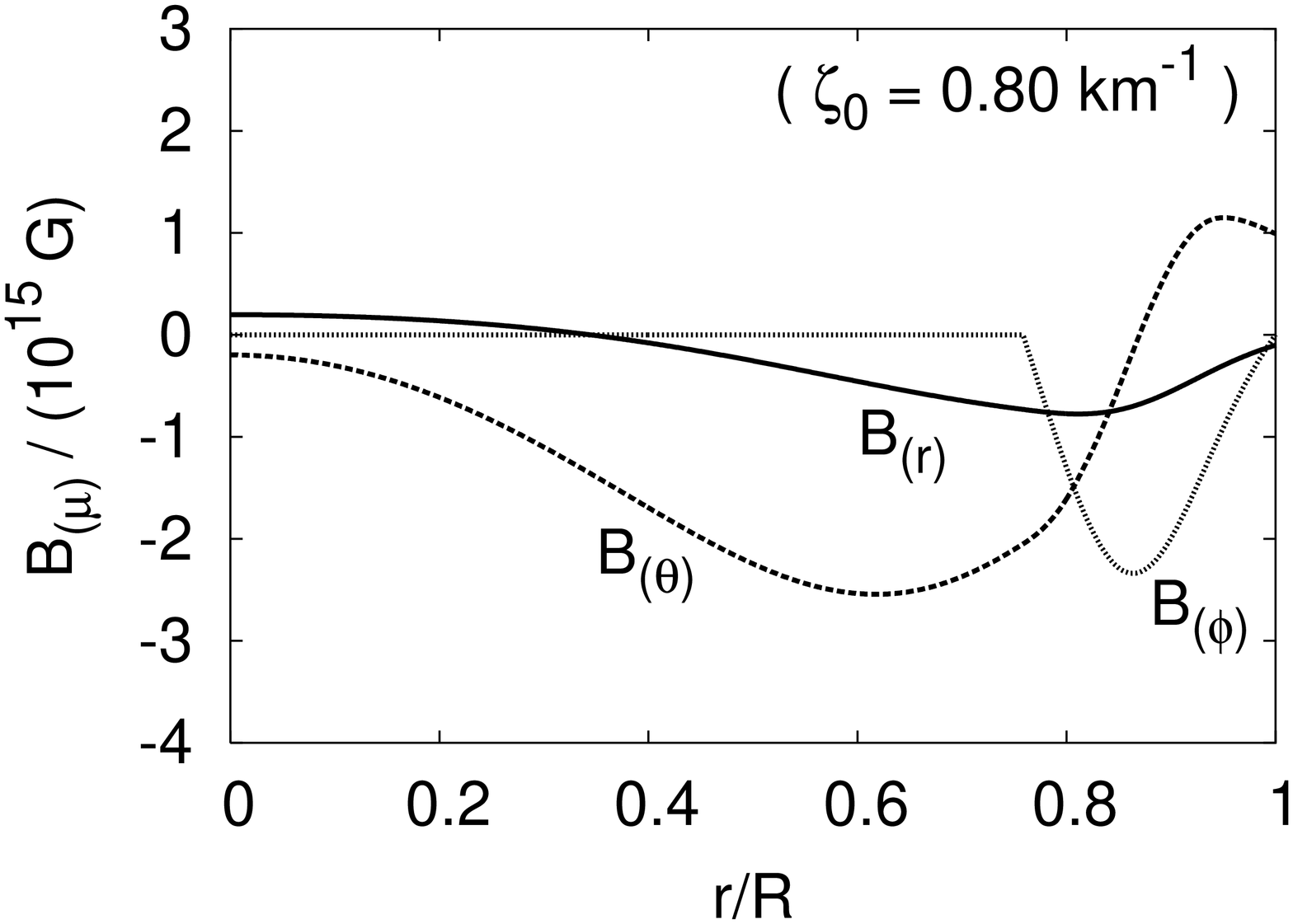}
\end{center}
\caption{ The profiles of the tetrad components of the magnetic field
($B_{(r)}(\theta=0)$, $B_{(\theta)}(\theta=\pi/2)$,
$B_{(\phi)}(\theta=\pi/2)$) are shown for $\zeta_0=0$ km$^{-1}$,
$\zeta_0=0.40$ km$^{-1}$ and $\zeta_0=0.80$ km$^{-1}$, and $l=1,3$.
\label{fig6}}
\end{minipage}
\end{figure*}
\begin{figure*}
\centering
\begin{minipage}{176mm}
\begin{center}
\hskip 1.7cm
\includegraphics[width=5.9cm,angle=0]{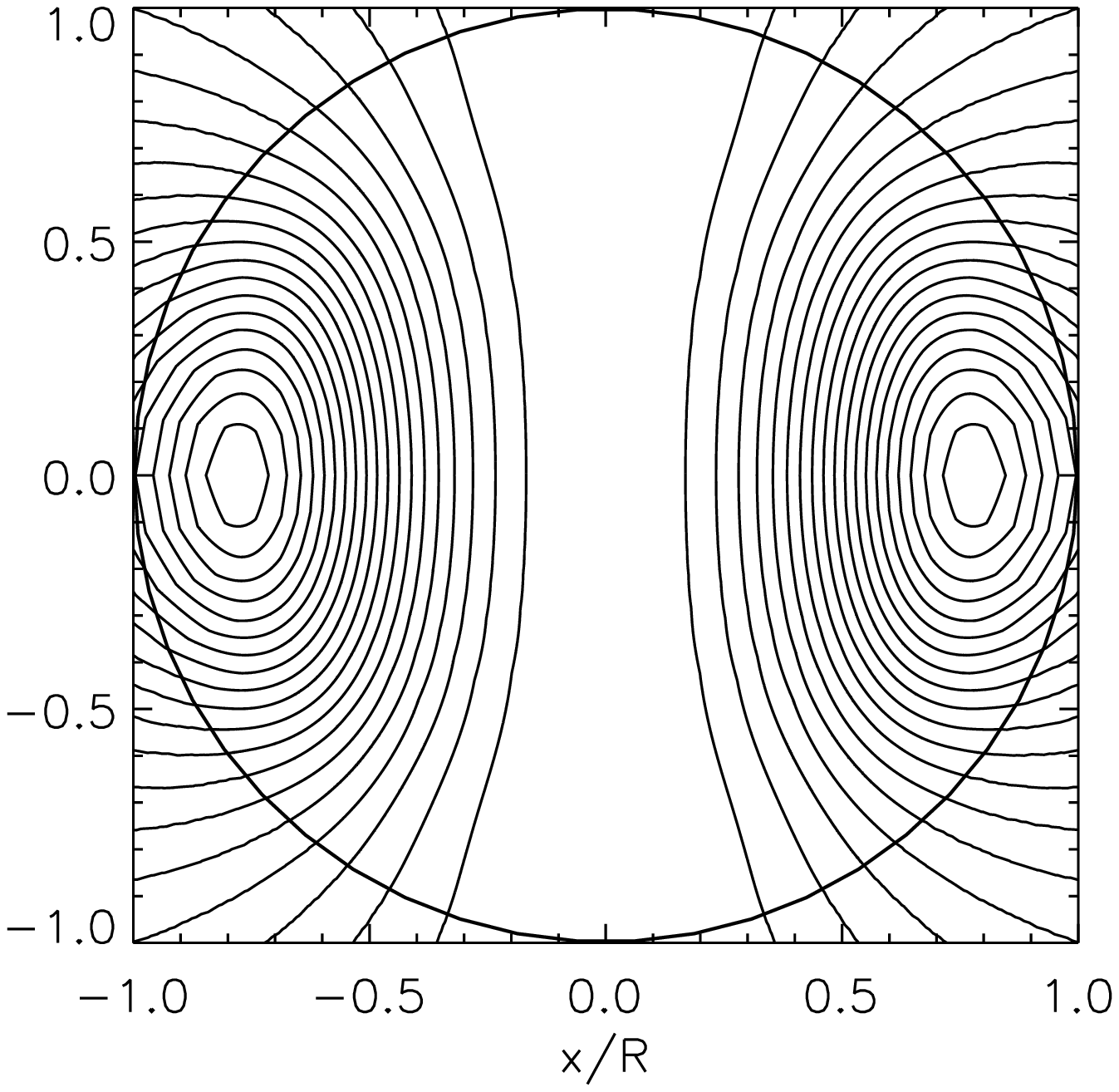}
\hskip -1cm
\includegraphics[width=5.9cm,angle=0]{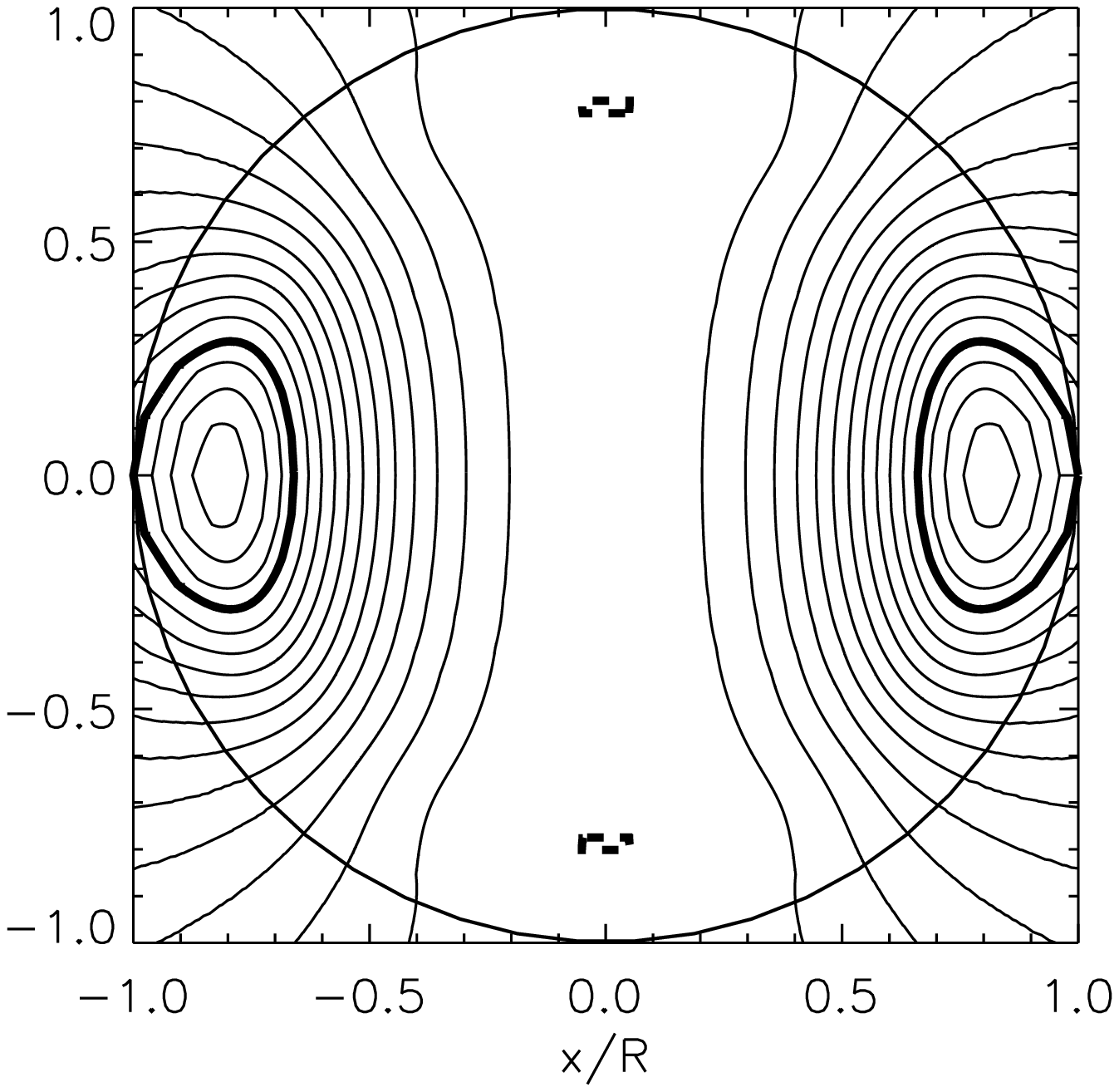}
\hskip -1cm
\includegraphics[width=5.9cm,angle=0]{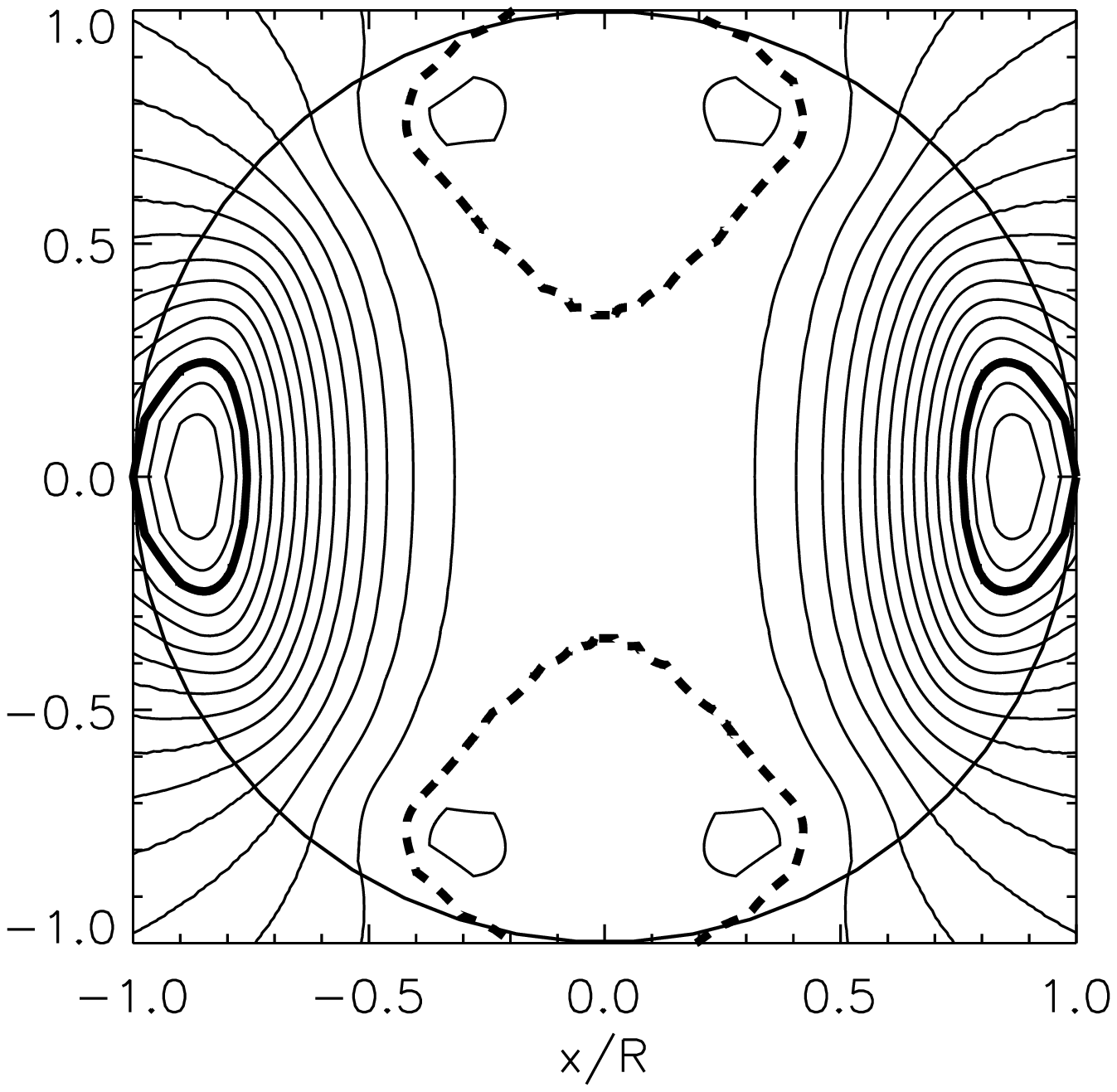}
\end{center}
\caption{The projection of the field lines in the meridional plane is
shown for $\zeta_0=0,0.40,0.80$ km$^{-1}$ respectively, and $l=1,3$.
The dashed lines correspond to the $\psi=0$ surfaces, and the toroidal
field is confined within the marked region.
\label{fig7}}
\end{minipage}
\end{figure*}
\begin{figure*}
\centering
\begin{minipage}{176mm}
\begin{center}
\hskip 1.7cm
\includegraphics[width=5.9cm,angle=0]{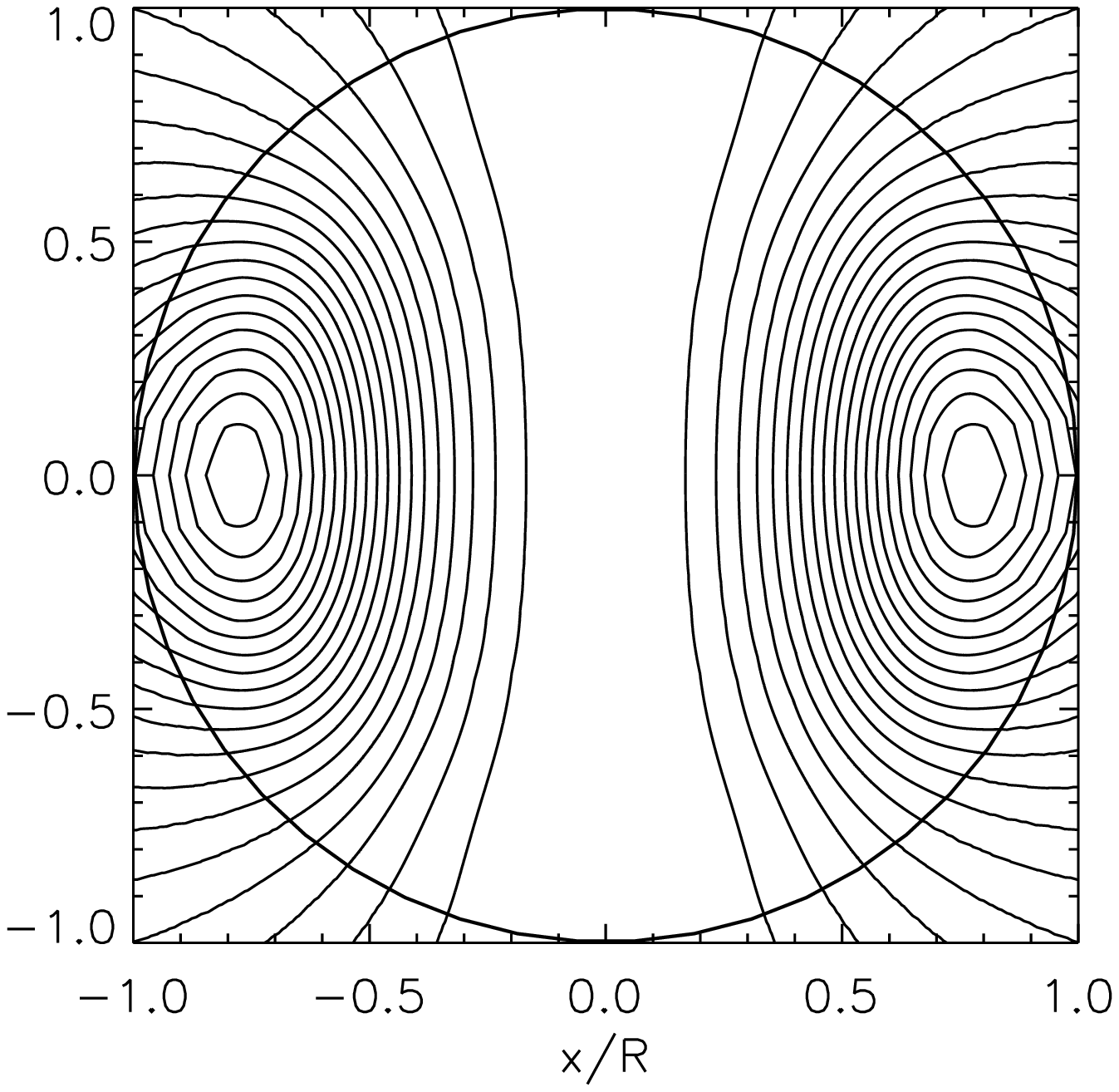}
\hskip -1cm
\includegraphics[width=5.9cm,angle=0]{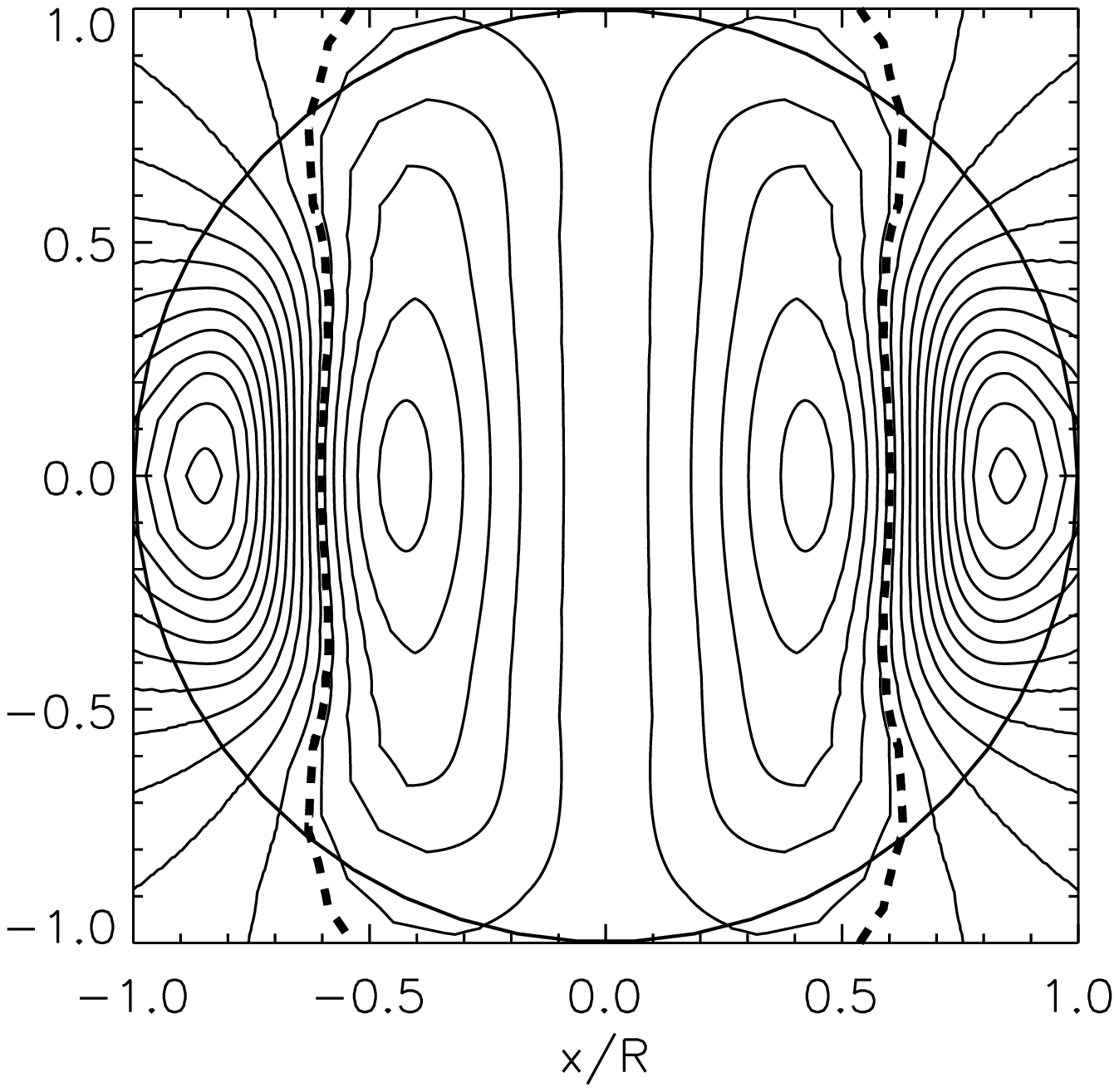}
\hskip -1cm
\includegraphics[width=5.9cm,angle=0]{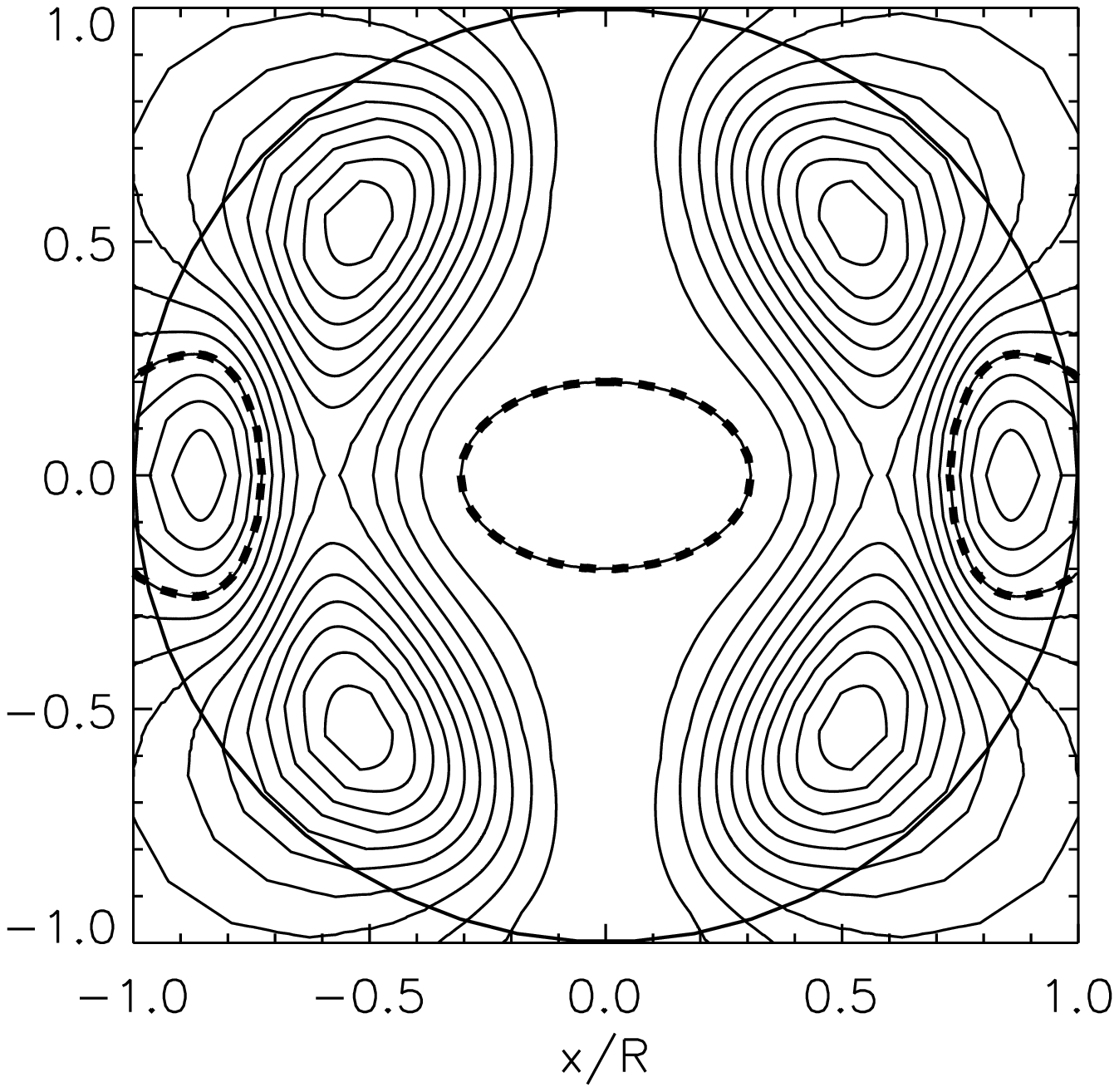}
\end{center}
\caption{The projection of the field lines in the meridional plane is
shown for $\zeta_0=0$ km$^{-1}$ and $l=1,3$.  The left panel refers to
the solution corresponding to the absolute minimum of
$|a_3(R)/a_1(R)|$; in this solution $\psi$ has no nodes. The center
and right panels refer to solutions corresponding to relative minima
of $|a_3(R)/a_1(R)|$; in these cases $\psi$ has one and two nodes,
respectively.  The dashed lines corresponds to the $\psi=0$ surfaces.
\label{fig10}}
\end{minipage}
\end{figure*}
\begin{figure*}
\centering
\begin{minipage}{176mm}
\begin{center}
\includegraphics[width=4cm,angle=270]{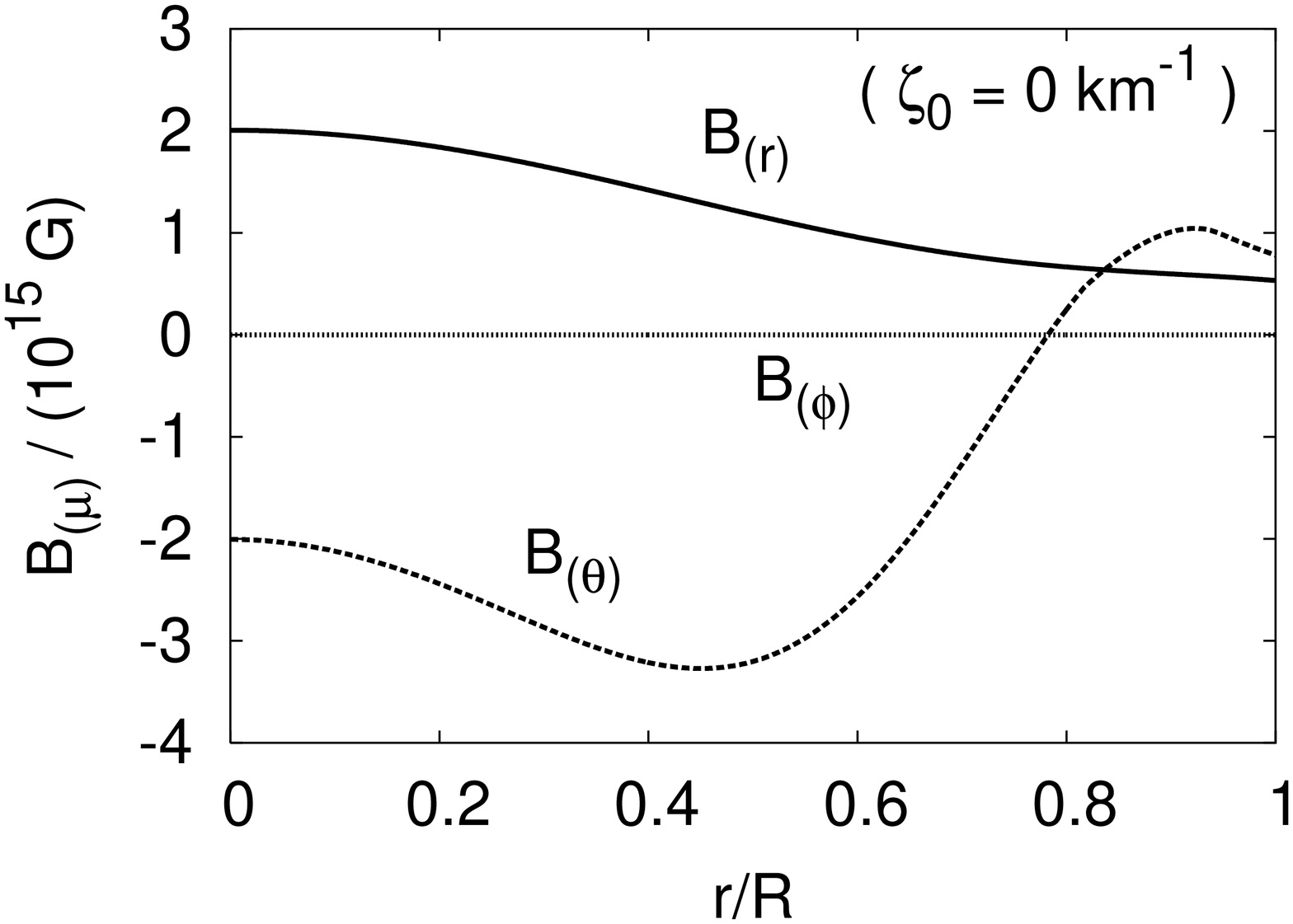}  
\includegraphics[width=4cm,angle=270]{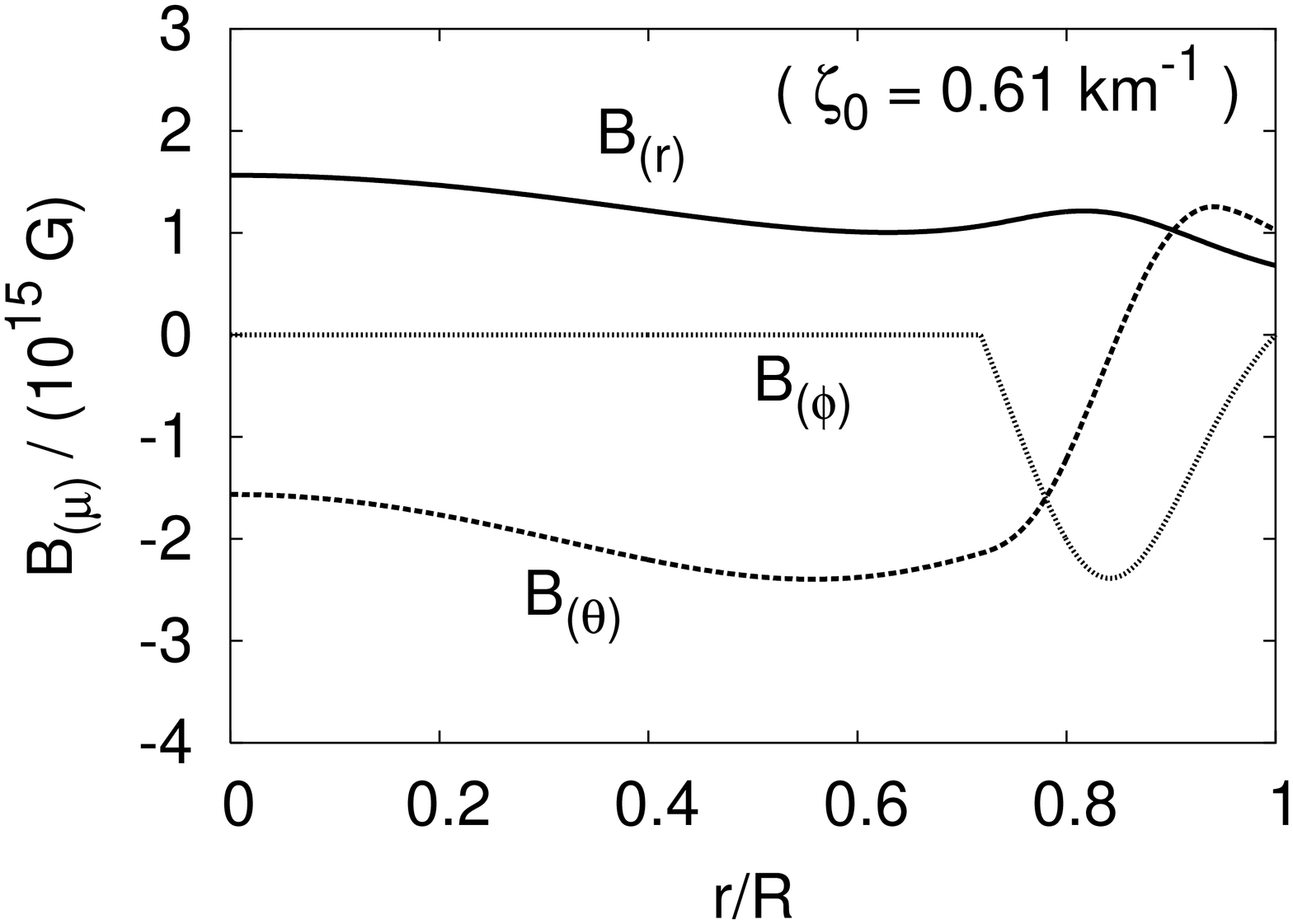}    
\includegraphics[width=4cm,angle=270]{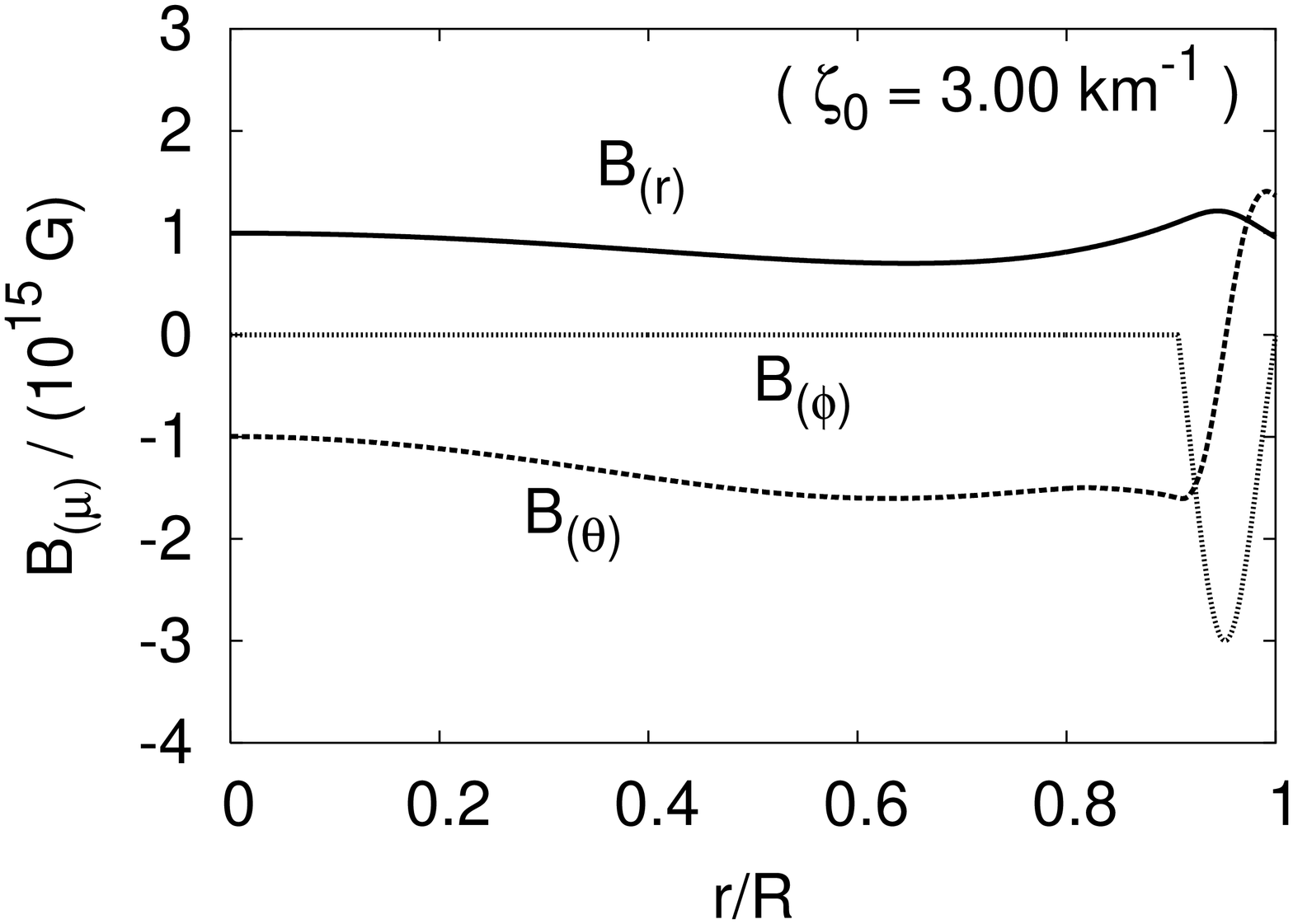}
\end{center}
\caption{The profiles of the tetrad components of the magnetic field
($B_{(r)}(\theta=0)$, $B_{(\theta)}(\theta=\pi/2)$,
$B_{(\phi)}(\theta=\pi/2)$) for the case including $l=1,3,5$, with
$\zeta_0=0$ km$^{-1}$, 
$\zeta_0=0.61$ km$^{-1}$ 
and $\zeta_0=3.00$ km$^{-1}$.
\label{fig11}}
\end{minipage}
\end{figure*}
\begin{figure*}
\centering
\begin{minipage}{176mm}
\begin{center}
\hskip 1.7cm
\includegraphics[width=5.9cm,angle=0]{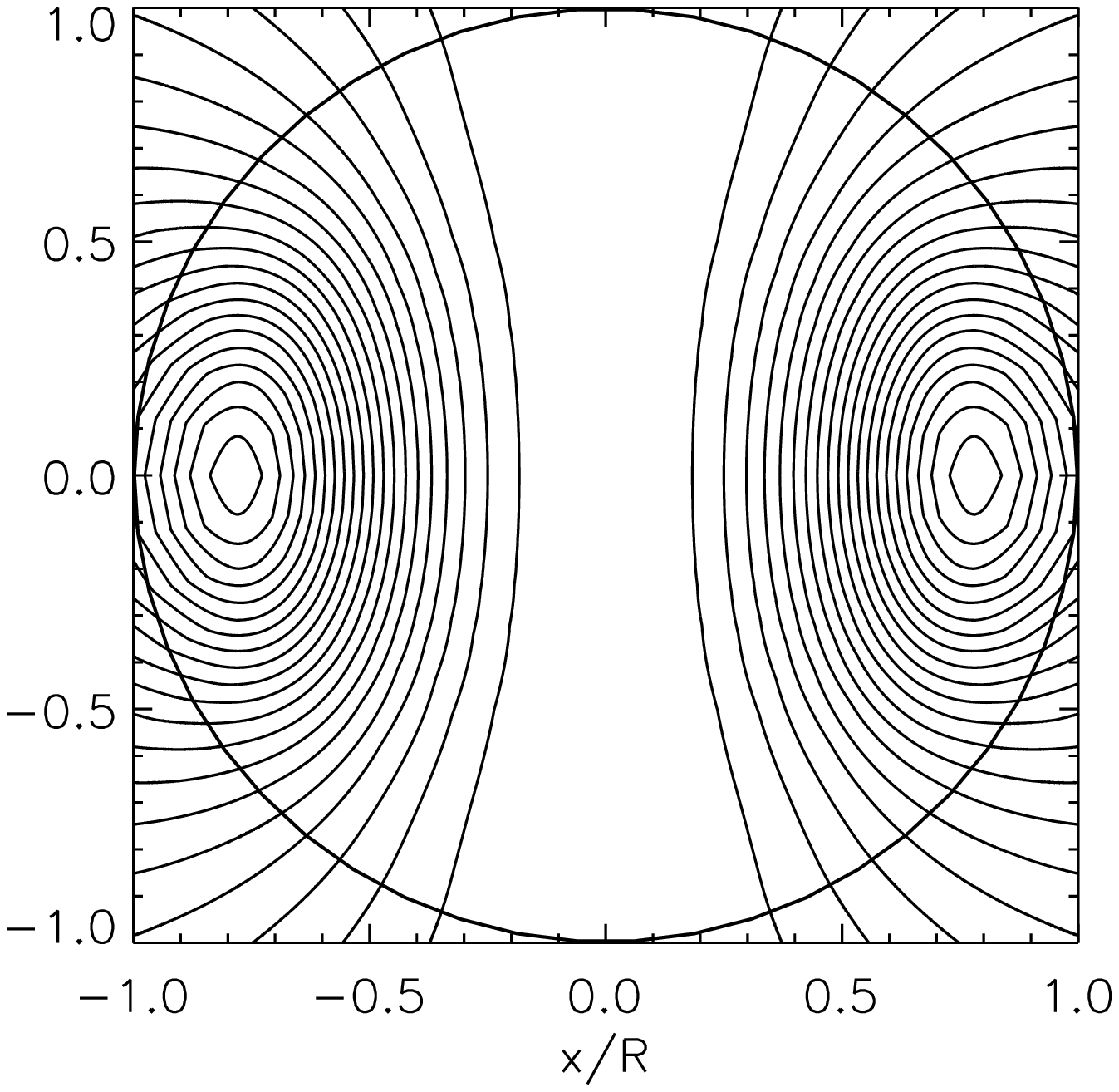}
\hskip -1cm
\includegraphics[width=5.9cm,angle=0]{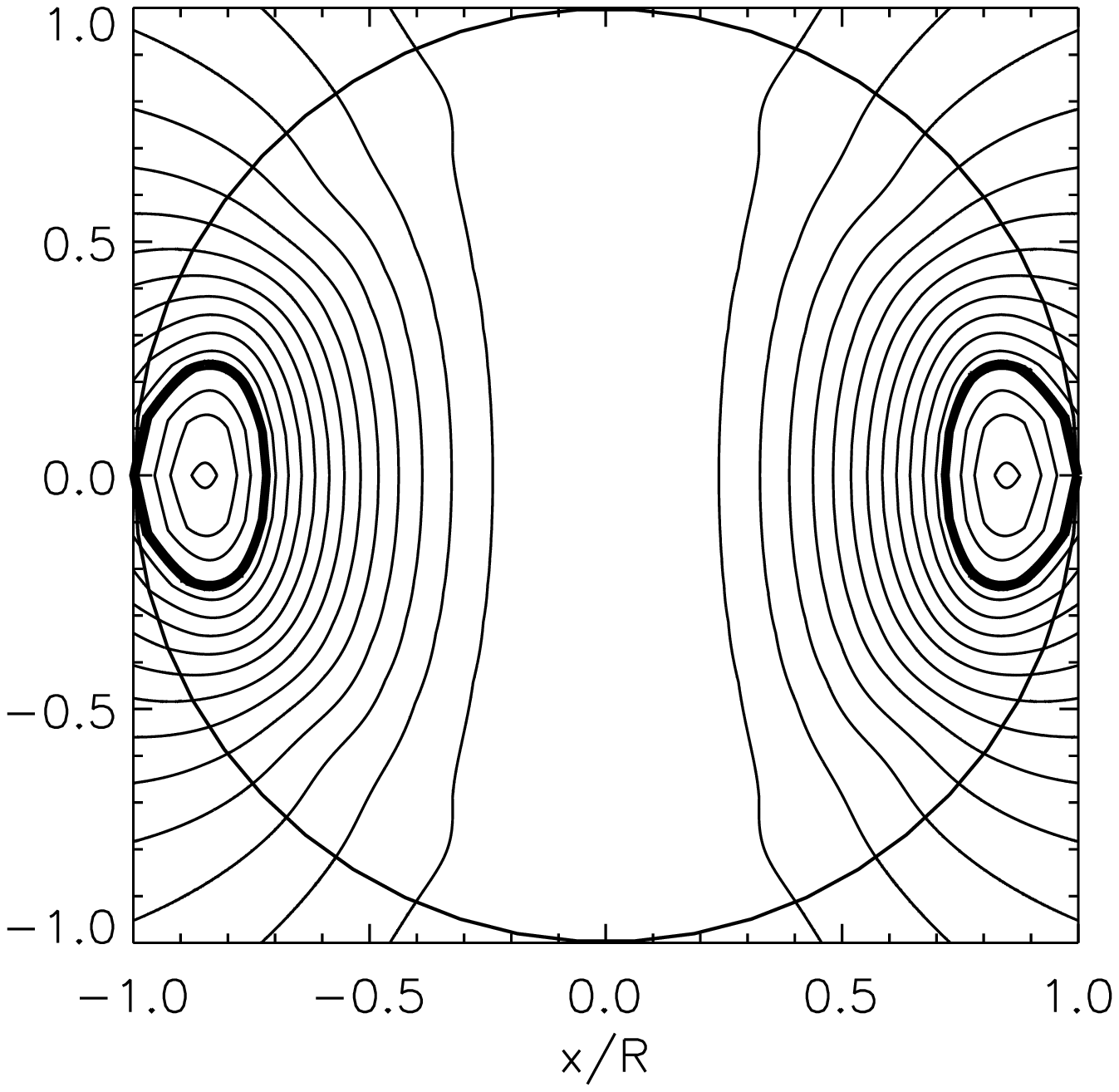}
\hskip -1cm
\includegraphics[width=5.9cm,angle=0]{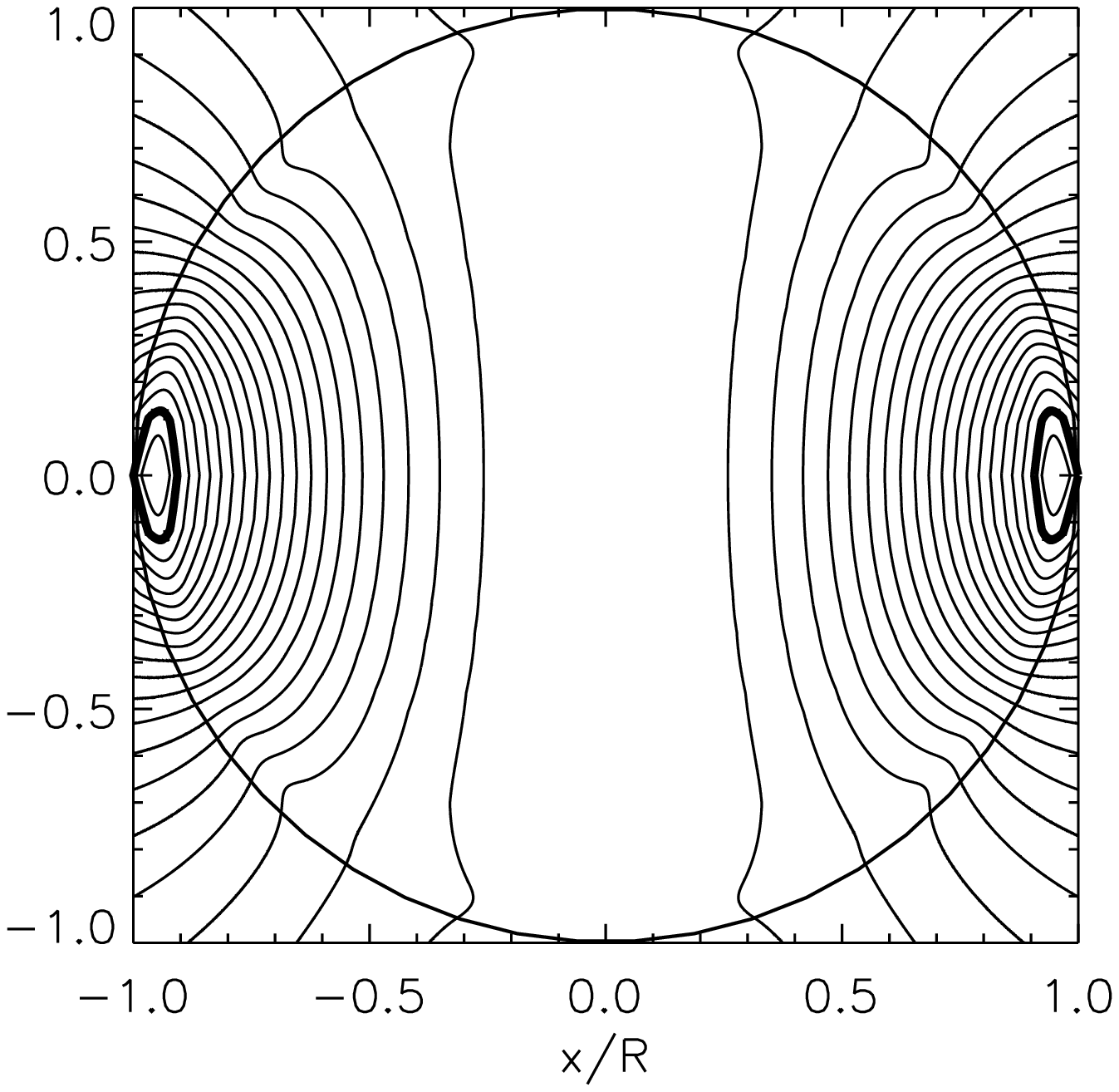}
\end{center}
\caption{The projection of the field lines in the meridional plane is
shown for $\zeta_0=0$ km$^{-1}$, $\zeta_0=0.61$ km$^{-1}$ and
$\zeta_0=3.00$ km$^{-1}$ respectively, and for $l=1,3,5$.  The
toroidal field is confined within the marked region.
\label{fig13}}
\end{minipage}
\end{figure*}

In Section \ref{L1L2}, we set the value of $a_1$ at the surface to be
$1.93\cdot 10^{-3}$ km and we minimized the $l=2$ contribution.  It is
clear that, since the $l=1$ and $l=2$ multipoles belong to different
families, any attempt to minimize the relative weight of one with
respect to the other leads to the trivial solution.  The properties of
equation (\ref{new22}) discussed above, tell us that if $a_1(R)\neq 0$
we cannot consistently set to zero the remaining odd order components
$a_{2l+1}$. However, we have the freedom of setting to zero all even
terms $a_{2l}$.  Therefore, since we have chosen to minimize the
contributions of the $l> 1$ harmonics outside the star, we shall focus
on the symmetric family of solutions ($a_{2l}\equiv0$); we will
briefly discuss an example belonging to the antisymmetric family in
subsection \ref{even-arm}.
\subsection{The case with $l=1$ and $l=3$}\label{L1L3}
We now consider the system of equations including only the $l=1$ and
$l=3$ components.  The projected system is
\begin{eqnarray}
&&\frac{1}{4\pi}\left(e^{-\lambda}a_1''
+e^{-\lambda}\frac{\nu'-\lambda'}{2}a_1'-\frac{2}{r^2}a_1 \right)
\nonumber\\
&&-\frac{e^{-\nu}}{4\pi} \int_0^\pi (3/4) 
\; \zeta_0^2 \left(\psi-3\psi |\psi/\bar{\psi}|
+2\psi^3/\bar{\psi}^2 \right)
\nonumber\\
&&\cdot\Theta(|\psi/\bar{\psi}|-1)\sin{\theta}\;d\theta\;\nonumber\\
&&=\left[c_0-\frac{4}{5}c_1\left(a_1-\frac{3}{7}a_3\right)\right] 
(\rho+P)r^2\,,  \label{new31}
\end{eqnarray}
\begin{eqnarray}
&&\frac{1}{4\pi}\left(e^{-\lambda}a_3''
+e^{-\lambda}\frac{\nu'-\lambda'}{2}a_3'-\frac{12}{r^2}a_3 \right)
\nonumber\\
&&+\frac{e^{-\nu}}{4\pi} \int_0^\pi (7/48) 
\; \zeta_0^2 \left(\psi-3\psi |\psi/\bar{\psi}|
+2\psi^3/\bar{\psi}^2 \right)
\nonumber\\
&&\cdot\Theta(|\psi/\bar{\psi}|-1)(3-15\cos^2{\theta})\sin{\theta}\;d\theta
\nonumber\\
&&=\frac{2}{15}c_1 (\rho+P)r^2 (a_1-4a_3)  \;\; ,
\label{new32}
\end{eqnarray}
where 
\begin{equation}
\psi=\left[-a_1+\frac{a_3 (3-15\cos^2{\theta})}{2}\right] 
\sin^2{\theta}
  \;\; .
\label{psi13}
\end{equation} 
We again impose regularity at the origin (Eq. (\ref{alphal})),
continuity in $r=R$ of $a_1,a_1',a_3,a_3'$ with the vacuum solutions
for $a_1(r)$, $a_3(r)$ given by Eq. (\ref{vacsol}), and we fix
$a_1(R)=1.93\cdot 10^{-3}$ km by normalization.  For the remaining
constraint we choose the solution that minimizes the absolute value of
$a_3(R)$. We find that there is a discrete series of local minima of
$|a_3(R)|$, and we select among them the absolute minimum.

Fig. \ref{fig6} shows the profiles of the tetrad field components
(see Eq. (\ref{new24a})) obtained by numerically integrating
Eqns. (\ref{new31}), (\ref{new32}), for different values of $\zeta_0$.
$B_{(r)}$ is evaluated at $\theta=0$, while $B_{(\theta)}$,
$B_{(\phi)}$ are evaluated at $\theta=\pi/2$.  As $\zeta_0$ increases,
the magnitude of the toroidal field becomes larger, but the region
where it is confined shrinks, as already found in section
\ref{dipole}.  The projection of the field lines in the meridional
plane is shown in Fig. \ref{fig7} for the same values of $\zeta_0$.
It shows that, for $\zeta_0\gtrsim0.40$ km$^{-1}$, the magnetic field
lines lie in disconnected regions, separated by dashed lines in the
figure.  Inside these regions, the function $\psi$ has opposite sign
and no toroidal field is present. A similar phenomenon has been
discussed in Colaiuda {\it et al.}  \shortcite{colaiuda2008}.  As we
will see in the next Section, the occurrence of these regions is an
artifact of the truncation in the harmonic expansion, and disappears
as higher order harmonics are included.

For completeness we also mention that the solutions corresponding to
the local minima of $|a_3(R)|$ different from the absolute minimum,
correspond to very peculiar field configurations (see
Fig. \ref{fig10}). The function $\psi$ has nodes on the equatorial
plane, therefore the field lines lie in disconnected regions; for a
fixed value of $\zeta_0$, the number of nodes increases as $|a_3(R)|$
increases.  These peculiar solutions exist for any value of $\zeta_0$,
and appear also when higher order harmonic components are
considered. Thus, they are not artifacts of the $l$-truncation.
\begin{figure*}
\centering
\begin{minipage}{176mm}
\begin{center}
\includegraphics[width=4.7cm,angle=270]{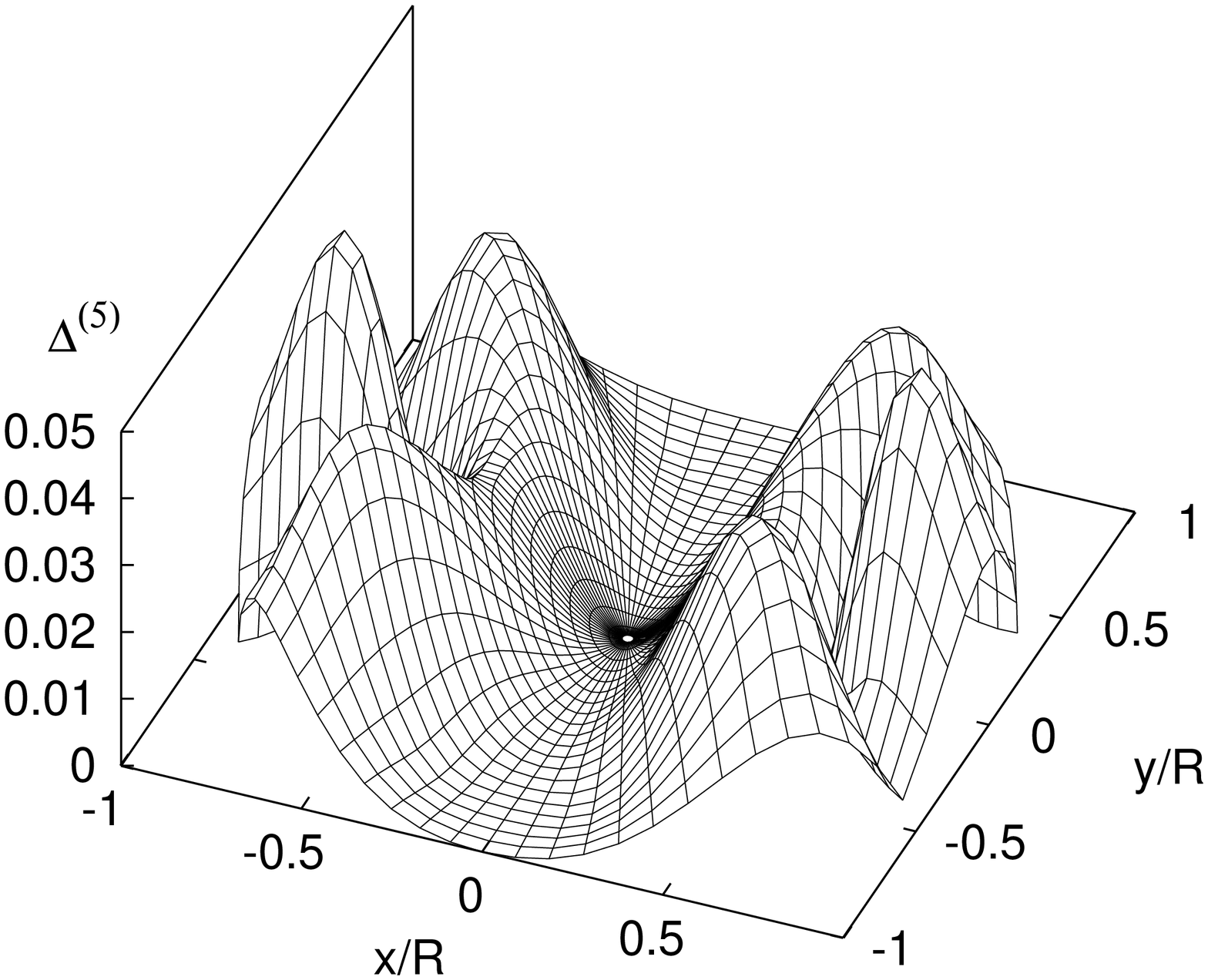}
\includegraphics[width=4.7cm,angle=270]{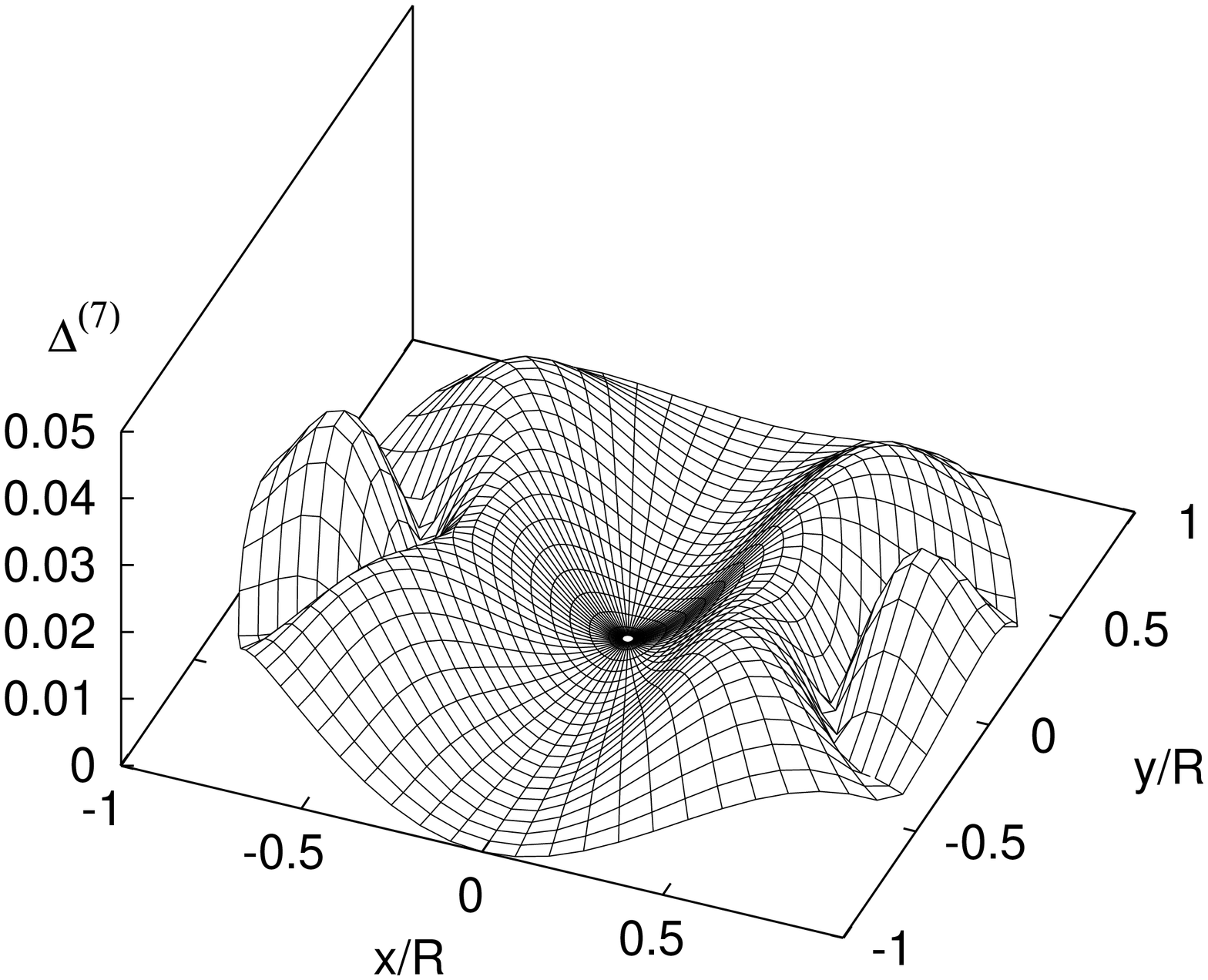}
\vskip 0.4cm
\includegraphics[width=4.7cm,angle=270]{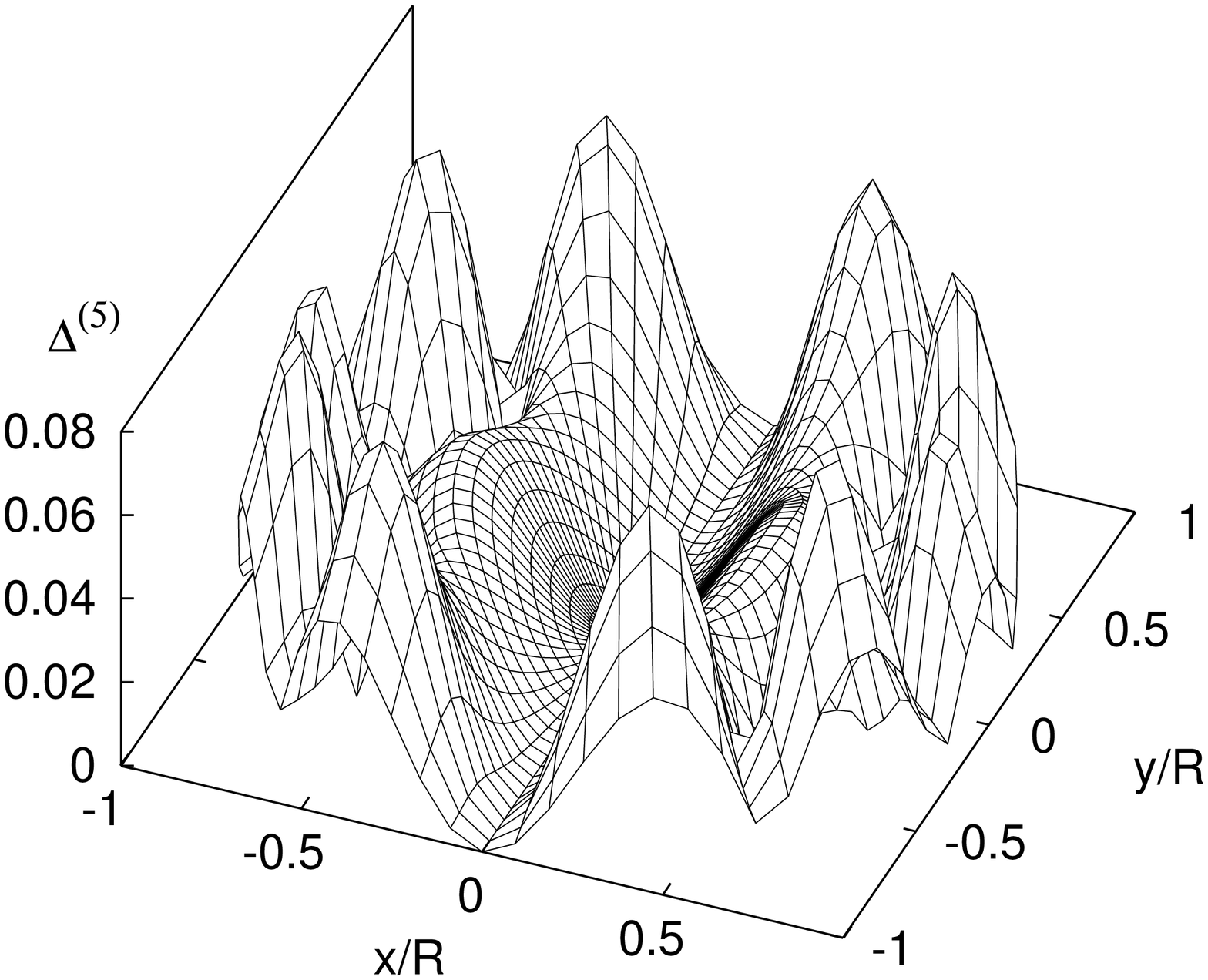}
\includegraphics[width=4.7cm,angle=270]{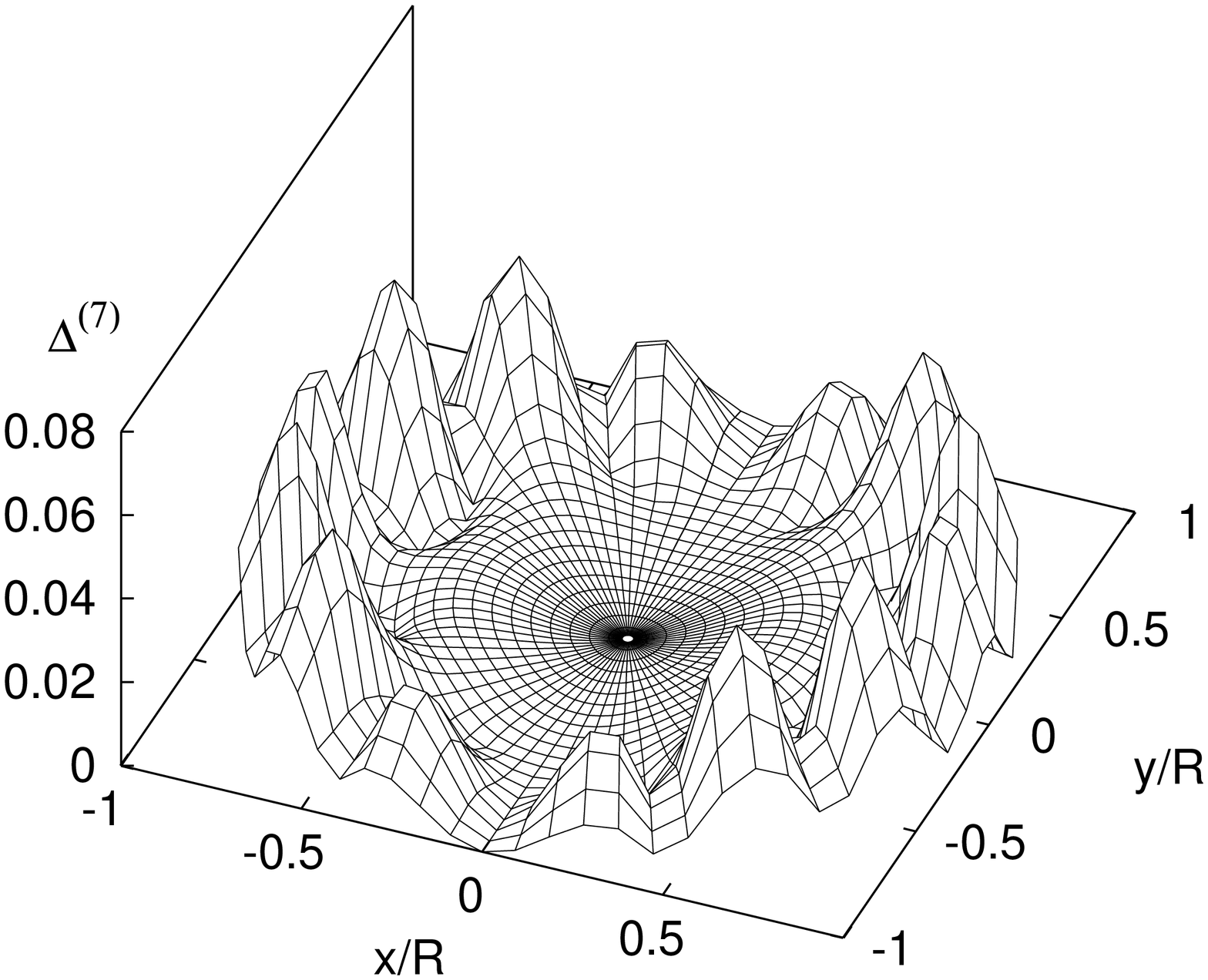}
\end{center}
\caption{The functions $\Delta^{(5)}$ (left panels) and $\Delta^{(7)}$
(right panels) are shown for $\zeta_0=0$ (upper panels) and
$\zeta_0=0.61$ km$^{-1}$ (lower panels) in the meridional plane for
$0\le r\le R$.\label{discr}}
\end{minipage}
\end{figure*}
\begin{figure*}
\centering
\begin{minipage}{176mm}
\begin{center}
\hskip 1.7cm
\includegraphics[width=6cm,angle=0]{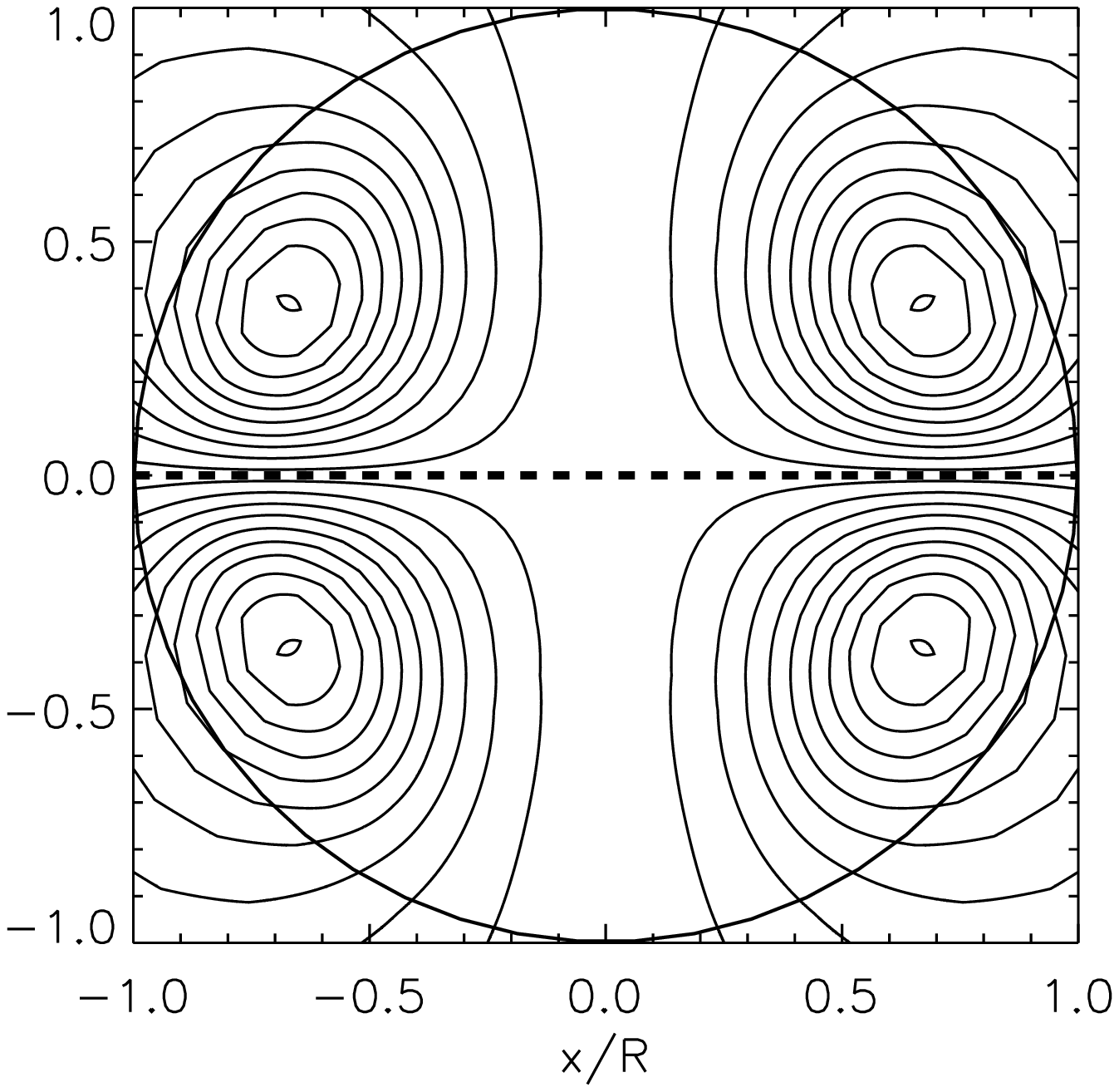} 
\includegraphics[width=6cm,angle=0]{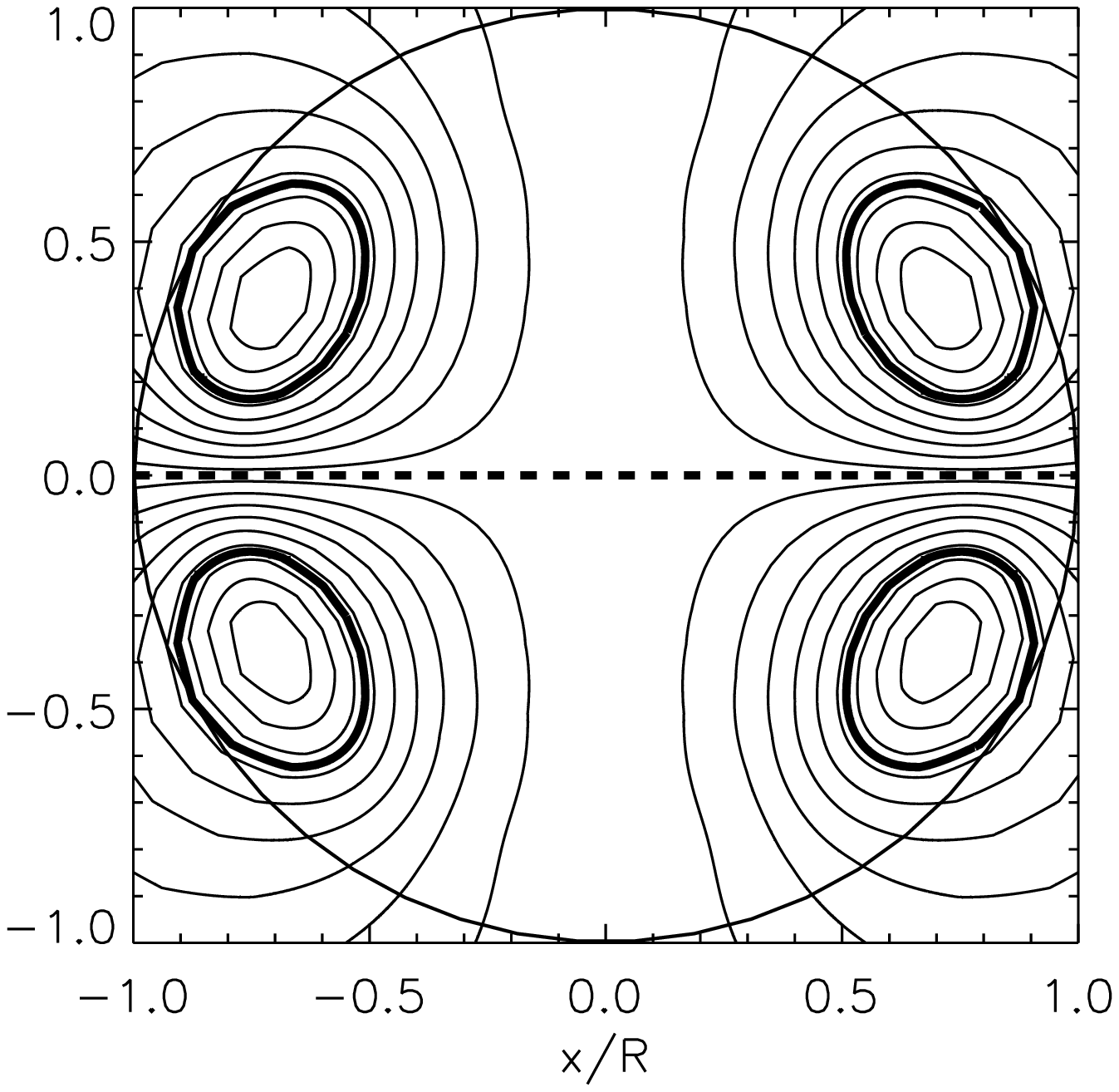} 
\end{center}
\caption{The projection of the field lines in the meridional plane is
shown for $\zeta_0=0$ km$^{-1}$ and $\zeta_0=0.30$ km$^{-1}$
respectively, and for $l=2,4$.  The dashed line corresponds to
$\psi=0$; the toroidal field is confined within the marked region.
\label{evenfig}}
\end{minipage}
\end{figure*}
\subsection{The case with $l=1,3,5$}\label{L1L3L5}
We now include the $l=5$ contribution. The three equations obtained by
projecting the GS equation (\ref{new22}) onto $l=1,3,5$ are given in
the Appendix \ref{eql5}. The boundary conditions are essentially the
same as in the previous Section; in particular, we look for the
absolute minimum of $a_3(R)^2+a_5(R)^2$, with fixed $a_1(R)=1.93\cdot
10^{-3}$ km.

In Fig. \ref{fig11} the profiles of the tetrad components of the
magnetic field are plotted for values of $\zeta_0$ in the range
$0\le\zeta_0\le3.00$ km$^{-1}$. Fig. \ref{fig13} shows the projections
of the field lines in the meridional plane corresponding to the same
values of $\zeta_0$.  Comparing the results with the case $l=1,3$ we
see that the presence of the harmonic $l=5$ modifies the magnetic
field shape, but most of the features discussed in the previous
Section are still present.

An interesting difference is the following. While in the case $l=1,3$
for $\zeta_0\gtrsim0.40$ km$^{-1}$ we find field configurations which
exhibit two disconnected regions where the function $\psi$ has
opposite sign and the magnetic field lines are confined (regions
within dashed lines in Fig. \ref{fig7}), this does not occur when the
$l=5$ component is taken into account.  This shows that the above
feature has to be considered as an artifact of the truncation in the
harmonic expansion.

\subsection{Higher order multipoles}\label{L1L3L5L7}

Up to now we have included components with $l<7$, neglecting the
contribution from $l\ge7$. In order to test the accuracy of this
approximation, we have studied the convergence of the harmonic
expansion. To this purpose, we have solved the GS equation
(\ref{new22}) including odd harmonic components up to $l=7$, for
$\zeta_0=0$ and $\zeta_0=0.61$ km$^{-1}$, and we have computed the
quantities
\begin{eqnarray}
\Delta^{(5)}(r,\theta)&=&\left|\frac{\psi_{l\le5}(r,\theta)-
\psi_{l\le3}(r,\theta)}{\bar\psi}\right|  \;\; ,
\nonumber\\
\Delta^{(7)}(r,\theta)&=&\left|\frac{\psi_{l\le7}(r,\theta)-
\psi_{l\le5}(r,\theta)}{\bar\psi}\right|  \;\; .
\end{eqnarray}
These functions are shown in Fig. \ref{discr}. They are plotted only
inside the star since outside they are much smaller.  Fig. \ref{discr}
shows that the error in neglecting $l\ge7$, quantified by the function
$\Delta^{(7)}$, is $\lesssim2\%$ for $\zeta_0=0$ and $\lesssim4\%$ for
$\zeta_0=0.61$ km$^{-1}$. Furthermore, a comparison of $\Delta^{(5)}$
and $\Delta^{(7)}$ shows that the harmonic expansion converges.

\subsection{An example of antisymmetric solution}\label{even-arm}
Here we show an example of a solution belonging to the antisymmetric
family corresponding to $l=2,4$. In Fig. \ref{evenfig} we plot the
field lines projected on the meridional plane, for $\zeta_0=0$
km$^{-1}$ and $\zeta_0=0.30$ km$^{-1}$.  We remark that the field
lines are antisymmetric with respect to the equatorial plane; as a
consequence, the total magnetic helicity is zero (see Section
\ref{helicity}).  Similar zero-helicity configurations have been
considered in Braithwaite \shortcite{brait-nonax}.
\section{Magnetic helicity and energy}\label{helicity}
The stationary configurations of magnetized neutron stars which we
have found, depend on the value of the free parameter $\zeta_0$,
i.e. on the ratio between the toroidal and the poloidal components of
the magnetic field.  In this Section, we provide an argument to assign
a value to $\zeta_0$.  Furthermore, we compute the magnetic energy of
the system to compare the contributions from poloidal and toroidal
fields.

The total energy of the system (the star, the magnetic field and the
gravitational field) can be determined by looking at the far field
limit ($r\rightarrow\infty$) of the spacetime metric
\cite{MTW,Thorne}. Following Colaiuda {\it et al.}
\shortcite{colaiuda2008}, Ioka \& Sasaki \shortcite{IS}, we write the
perturbed metric as
\begin{eqnarray}
ds^2&=&-e^{\nu}\Big[ 1+2h(r,\theta) \Big] dt^2
+e^{\lambda}\biggl[1+\frac{2e^{\lambda}}{r}m(r,\theta)\biggr]dr^2
\nonumber\\
&&+r^2\Big[1+2 k(r,\theta)\Big]\biggl(d\theta^2+sin^2 \theta \,d\phi^2\biggr)
\nonumber\\
&&+2i(r,\theta) dtdr+2v(r,\theta) dtd\phi+2w(r,\theta) drd\phi
  \;\; 
\label{defmetric}
\end{eqnarray}
where, in particular, $m(r,\theta)=\sum_lm_l(r)P_l(\cos{\theta})$. The total
mass-energy of the system is
\begin{equation}
E=M+\delta M
\end{equation}
where $M$ is the gravitational mass of the unperturbed star and
\begin{equation}
\delta M=\lim_{r\rightarrow\infty}m_0(r)\,.
\end{equation}
In Appendix \ref{energy}, we discuss the equations which allow to
determine $E$.  We remark that $\delta M$ includes different
contributions, due to magnetic energy, deformation energy, etc.

In order to evaluate the magnetic contribution to $E$, it is
convenient to use the Komar-Tolman formula (see for instance Straumann
\shortcite{Straumann}, Chap. 4) for the total energy:
\begin{equation}
E=2\int_V\left(T_{\mu\nu}-\frac{1}{2}Tg_{\mu\nu}\right)\eta^\mu n^\nu dV
\end{equation}
(where $V$ is the $3$-surface at constant time, $\eta^\mu$ is the
timelike Killing vector, $n^\mu$ is the normalized, future directed
normal to $V$); the magnetic contribution comes from the stress-energy
tensor of the electromagnetic field, $T_{em}^{\mu\nu}$ (\ref{new8}),
i.e.
\begin{eqnarray}
E_m&=&2\int_V\left(T^{em}_{\mu\nu}-\frac{1}{2}T^{em}g_{\mu\nu}\right)
\eta^\mu n^\nu dV\nonumber\\
&=&\frac{1}{2} \int_0^{\infty}
r^2 e^{\frac{\lambda+\nu}{2}} dr
\int_0^\pi \sin{\theta} \;B^2 d\theta \,.
\end{eqnarray}

The total (integrated) magnetic helicity $H_m$ of the field
configuration is
\begin{equation}
H_m=\int d^3 x \sqrt{-g} H^0_m  \;\; ,
\label{EH1}
\end{equation}
where $H_m^0$ is the $t$-component of the magnetic helicity 4-current,
defined as
\begin{equation}
H^\alpha_m=\frac{1}{2}\epsilon^{\alpha\beta\gamma\delta}F_{\gamma\delta}A_\beta
  \;\; .
\label{EH3}
\end{equation}
Explicitly, we have
\begin{equation}
H_m=-2\pi\int_0^R dr \int_0^\pi 
[A_r \psi_{,\theta}-\psi A_{r,\theta}] d\theta
  \;\; , 
\label{EH6}
\end{equation}
where
\begin{eqnarray}
\psi A_{r,\theta}&=& \frac{e^{\frac{\lambda-\nu}{2}}}{\sin{\theta}} 
\psi^2 \zeta_0 \left( |\psi/\bar{\psi}|-1 \right)
\cdot\Theta(|\psi/\bar{\psi}|-1)
  \;\; ,\nonumber\\
\psi_{,\theta} A_r&=& \psi_{,\theta} e^{\frac{\lambda-\nu}{2}}\zeta_0 
\int_0^\theta \frac{\psi}{\sin{\theta'}}
\left( |\psi/\bar{\psi}|-1 \right)
\nonumber\\
&&\qquad\qquad\qquad\qquad\cdot\Theta(|\psi/\bar{\psi}|-1) d\theta'
   \;\; .
\label{EH8}
\end{eqnarray}
The functional dependence of $H_m$ on the potential of the toroidal
field, $A_r$ (see equation (\ref{EH6})), shows that regions of space
where the toroidal field vanishes do not contribute to the magnetic
helicity.

In ideal MHD, the magnetic helicity is a conserved quantity
\cite{bek,BS}. Thus, if we consider magnetic field configurations
having the same value of the magnetic helicity and different energies,
the lowest energy configuration is energetically favoured.

\begin{figure}
\begin{center}
\includegraphics[width=5cm,angle=-90]{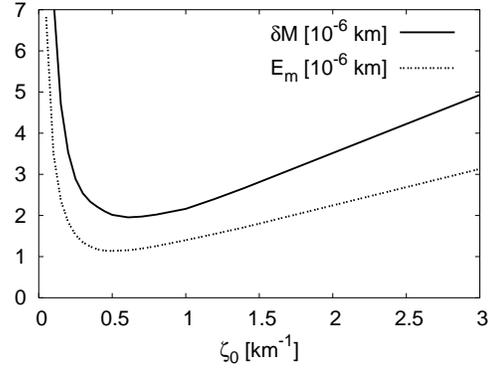} 
\end{center}
\caption{The functions $\delta M$ and $E_m$ are plotted as 
functions of $\zeta_0$, for $l=1,3,5$ and $H_m=1.75 \cdot 10^{-6}$ km$^2$.
\label{EHfig1}}
\end{figure}

In Fig. \ref{EHfig1} we plot $\delta M$ and $E_m$ as functions of
$\zeta_0$, for a fixed helicity $H_m=1.75 \cdot 10^{-6}$ km$^2$.
$\delta M$, and consequently the total energy, has a minimum at
$\zeta_0=0.61$ km$^{-1}$.  A fixed value of $H_m$ corresponds, for any
assigned value of $\zeta_0$, to a different normalization constant
$B_{pole}$.  Since $\delta M$, $H_m$ and $E_m$ have the same quadratic
dependence on the magnetic field normalization, this means that if we
change $H_m$ the plots of $\delta M$ and of $E_m$ as functions of
$\zeta_0$ are simply rescaled with respect to that shown in Fig.
\ref{EHfig1}.  Consequently, for any fixed value of $H_m$ the position
of the minimum of the total energy is the same as that shown in Fig.
\ref{EHfig1}.  We conclude that the configuration with
$\zeta_0\simeq0.61$ km$^{-1}$ is energetically favoured. This
configuration is shown, among others, in Figs. \ref{fig11},
\ref{fig13}.  From Fig. \ref{EHfig1} we also see that the contribution
of the magnetic energy to $\delta M$ is of the order
$\sim50\%$-$70\%$.

\begin{figure}
\begin{center}
\includegraphics[width=5cm,angle=-90]{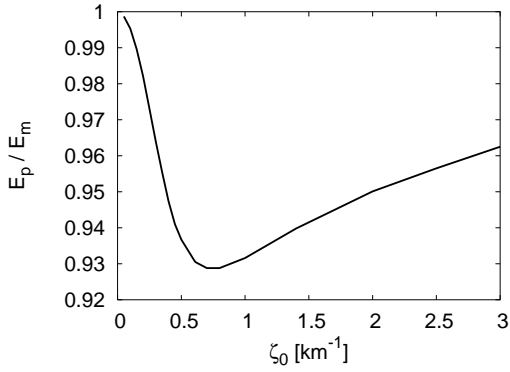}
\end{center}
\caption{The ratio $E_p/E_m$ is shown as a function of $\zeta_0$,
for $l=1,3,5$.\label{EPEM}}
\end{figure}

In Fig. \ref{EPEM} we show the ratio of poloidal to total magnetic
field energy, $E_p/E_m$, as a function of $\zeta_0$, for the
configurations ($l=1,3,5$) studied in this paper.  This plot is
interesting because the relative weight of the poloidal and the
toroidal components of the field significantly affects many
astrophysical processes involving magnetars, like magnetar activity
\cite{WT}, their thermal evolution \cite{PMG09}, their gravitational
wave emission \cite{Cutler}.  It should be mentioned that the surface
poloidal field is inferred from spin-down measurements which, however,
provide no hint about the toroidal field hidden inside the star.  We
find that for $\zeta_0=0.61$ km$^{-1}$, $E_p/E_m\simeq0.93$.

In a recent paper \cite{brait08} the stability of magnetic field
configurations of compact stars has been studied in the context of
Newtonian gravity, and assuming a polytropic equation of state. It has
been found that axisymmetric configurations are stable when
$0.01\lesssim E_p/E_m\lesssim 0.8$. Our configurations are outside
this range, i.e. $E_p/E_m>0.9$.  This difference may be attributed to
several reasons: our models are computed in the framework of general
relativity, we use a more realistic equation of state, we choose a
particular function $\zeta(\psi)$, which is linear in $\left(
|\psi/\bar{\psi}|-1 \right)$ (see Eq. (\ref{new13})). A different
power-law dependence may lead to a different contribution of the
toroidal field.  In any case, we observe a tendency in favour of
models with predominant poloidal fields when using arguments of
minimum energy, and we do not think that other functional forms of
$\zeta$ may result in configurations with most of the energy stored in
the toroidal field.  This issue deserves further investigations.

\section{Discussion and conclusions}\label{conclusions}

In this paper we find a twisted-torus family of solutions in the
framework of general relativity. The toroidal component of the
magnetic field vanishes outside the star: neither discontinuities
(associated to surface currents) nor the vanishing of the total
magnetic field outside the star have been imposed; this is an
improvement with respect to previous works
\cite{IS,colaiuda2008,Haskell}. It should be stressed that
twisted-torus configurations have been found to emerge as a final
outcome of Newtonian MHD simulations with generic initial conditions
\cite{BS0,BN,BS}.

In order to have a twisted-torus configuration, there must be a
non-linear relation between toroidal and poloidal fields, leading to
couplings between different multipoles. We have investigated the
contributions of different harmonics, and we have constructed
equilibrium configurations with $1\le l\le5$.  In order to fix the
boundary conditions, we imposed that outside the star the dipolar
component dominates, and minimized the $l>1$ contributions which,
however, remain non negligible.  We find that there exist two
particular, independent classes of solutions: those {\it symmetric}
(with respect to the equatorial plane), in which all even order
components vanish ($a_{2l}\equiv0$), and the {\it antisymmetric}
solutions, characterized by the vanishing of odd components
($a_{2l+1}\equiv0$). The latter have zero helicity by definition,
therefore any solution minimizing energy at fixed helicity has a
vanishing antisymmetric component.

Our models also depend on a parameter, $\zeta_0$, which determines the
ratio between toroidal and poloidal fields ( $\approx \zeta_0 R$), and
the length-scale of the region where the toroidal field is confined
($\propto \zeta_0^{-1}$). As $\zeta_0$ increases, the amplitude of the
toroidal field grows, but the region where it is confined shrinks.
This parameter can be estimated by minimizing the total energy at
fixed magnetic helicity. We find that, for our neutron star model
(equation of state APR2, $M=1.4~M_\odot$), this minimum occurs at
$\zeta_0=0.61$ km$^{-1}$.  Therefore, we expect that in the early
evolution of a strongly magnetized (fluid) neutron star, the natural
final outcome after MHD equilibrium is established, are twisted-torus
configurations with geometries similar to our solutions.

Finally we have computed the magnetic energy associated to the
poloidal and toroidal fields.  We find that, although the amplitudes
of both fields are of the same order of magnitude, and the toroidal
field in the interior can be larger than the poloidal field at the
surface (for instance, it is 2-3 times larger if $\zeta_0=0.61$
km$^{-1}$), the contribution of the toroidal field to the total
magnetic energy is $\lesssim10\%$, because this field is non vanishing
only in a finite region of the star.  As mentioned in Section
\ref{helicity}, a different power-law dependence of the function
$\zeta(\psi)$ on $\left( |\psi/\bar{\psi}|-1 \right)$, may lead to a
different contribution of the toroidal field and we plan to
investigate this issue in a forthcoming paper.
 
\vskip .5cm Note: After our paper has been submitted and sent to the
arXiv, a paper has appeared \cite{LJ} where a model of magnetar with
twisted-torus magnetic field configuration is developed in the
Newtonian framework. The results of Lander \& Jones \shortcite{LJ} are
in agreement with ours, in that they find the toroidal field to be
bounded to less than 7\% of the total magnetic field.

\section*{Acknowledgments}\label{acknow}
This work was partially supported by the Spanish grant AYA
2007-67626-C03-02 and COMPSTAR, an ESF Research Networking Programme.
\appendix
\section{GS equation for the case with \MakeLowercase{$l=1,3,5$}}\label{eql5}
If we include the $l=1,3,5$ components, the GS equation (\ref{new22})
projected onto the harmonics $l=1,3,5$ gives the following system:
\begin{eqnarray}
&&\frac{1}{4\pi}\left(e^{-\lambda}a_1''
+e^{-\lambda}\frac{\nu'-\lambda'}{2}a_1'-\frac{2}{r^2}a_1 \right)
\nonumber\\
&&-\frac{e^{-\nu}}{4\pi} \int_0^\pi (3/4) 
\; \zeta_0^2 \left(\psi-3\psi |\psi/\bar{\psi}|
+2\psi^3/\bar{\psi}^2 \right)\nonumber\\
&&\cdot\Theta(|\psi/\bar{\psi}|-1)\sin{\theta}\;d\theta\nonumber\\
&&=\left[c_0-\frac{4}{5}c_1\left(a_1-\frac{3}{7}a_3\right)\right] 
(\rho+P)r^2\,,  \label{new135a}
\end{eqnarray}
\begin{eqnarray}
&&\frac{1}{4\pi}\left(e^{-\lambda}a_3''
+e^{-\lambda}\frac{\nu'-\lambda'}{2}a_3'-\frac{12}{r^2}a_3 \right)
\nonumber\\
&&+\frac{e^{-\nu}}{4\pi} \int_0^\pi (7/48) 
\; \zeta_0^2 \left(\psi-3\psi |\psi/\bar{\psi}|
+2\psi^3/\bar{\psi}^2 \right)\nonumber\\
&&\cdot \Theta(|\psi/\bar{\psi}|-1)(3-15\cos^2{\theta})
\sin{\theta}\;d\theta\nonumber\\
&&=c_1 (\rho+P)r^2 \left(\frac{2}{15}a_1-\frac{8}{15}a_3+
\frac{10}{33}a_5\right)  \;\; ,
\label{new135b}
\end{eqnarray}
\begin{eqnarray}
&&\frac{1}{4\pi}\left(e^{-\lambda}a_5''
+e^{-\lambda}\frac{\nu'-\lambda'}{2}a_5'-\frac{30}{r^2}a_5 \right)
\nonumber\\
&&+\frac{e^{-\nu}}{4\pi} \int_0^\pi (11/60) 
\; \zeta_0^2 \left(\psi-3\psi |\psi/\bar{\psi}|
+2\psi^3/\bar{\psi}^2 \right)\nonumber\\
&&\cdot\Theta(|\psi/\bar{\psi}|-1)
\frac{(-315\cos^4{\theta}+210\cos^2{\theta}-15)}{8}
\sin{\theta}\;d\theta\nonumber\\
&&
=c_1 (\rho+P)r^2 \left(\frac{4}{21}a_3-\frac{20}{39}a_5\right)  \;\; ,
\label{new135c}
\end{eqnarray}
where 
\begin{eqnarray}
\psi&=&\left[-a_1+\frac{a_3 (3-15\cos^2{\theta})}{2}\right.
\nonumber\\
&&\left.+\frac{a_5 (-315\cos^4{\theta}+210\cos^2{\theta}-15)}{8}\right] 
\sin^2{\theta}  \;\; .  
\nonumber
\end{eqnarray}

\section{The energy of the system}\label{energy}
The perturbation of the total energy of the system can be determined
from the far field limit of the spacetime metric \cite{MTW,Thorne,IS}:
\begin{equation}
\delta M=\lim_{r\rightarrow\infty}m_0(r)  \;\; ,
\label{dMm}
\end{equation}
where the perturbed metric is given by equation (\ref{defmetric}).
The functions $h(r,\theta)$ and $m(r,\theta)$ are 
\begin{eqnarray}
h(r,\theta)&=&\sum_lh_l(r)P_l(\cos{\theta})\quad,
\nonumber\\
m(r,\theta)&=&\sum_lm_l(r)P_l(\cos{\theta})\quad.
\end{eqnarray}
The perturbed Einstein equations ($[tt]$ and $[rr]$ components),
projected onto $l=0$, allow to determine the quantity $m_0(r)$:
\begin{eqnarray}
&&m'_0-4\pi r^2\frac{\rho'}{P'}\delta p_0=\nonumber\\
&&\frac{1}{3}(a'_1)^2 e^{-\lambda} +\frac{6}{7}(a'_3)^2
e^{-\lambda}+\frac{15}{11}(a'_5)^2 e^{-\lambda}
+\frac{2}{3r^2}a_1^2\nonumber\\
&&+\frac{72}{7r^2}a_3^2 +\frac{450}{11r^2}a_5^2\nonumber\\
&&+\frac{e^{-\nu}}{4}
\left[\int_0^\pi \zeta_0^2 \left( |\psi/\bar{\psi}|-1 \right)^2
\Theta(|\psi/\bar{\psi}|-1) \frac{\psi^2}{\sin{\theta}} d\theta\right]
\;\; , \nonumber\\ 
&&h'_0-e^{2\lambda}m_0\left(
\frac{1}{r^2}+8\pi P\right)-4\pi re^{\lambda}\delta p_0=\nonumber\\
&&\frac{1}{3r}(a'_1)^2+\frac{6}{7r}(a'_3)^2+\frac{15}{11r}(a'_5)^2
-\frac{2e^{\lambda}}{3r^3}a_1^2
\nonumber\\
&&-\frac{72e^{\lambda}}{7r^3}a_3^2-\frac{450e^{\lambda}}{11r^3}
a_5^2\nonumber\\
&&+\frac{e^{\lambda-\nu}}{4r}\left[
\int_0^\pi \zeta_0^2 \left( |\psi/\bar{\psi}|-1 \right)^2
\Theta(|\psi/\bar{\psi}|-1)\frac{\psi^2}{\sin{\theta}}d\theta \right]
\,,\nonumber\\\label{eq0mh}
\end{eqnarray}
where $\delta p_0$ is the $l=0$ component of the pressure perturbation
(and vanishes outside the star), and
$\psi=\sin{\theta}\sum_{l=1,3,5}a_l P_{l ,\theta}$.
Using the relation (arising from $T^{r\nu}_{\;\; ;\nu}=0$) 
\begin{eqnarray}
\delta p'_0 &=& -\frac{\nu'}{2}\left( \frac{\rho'}{P'}+1 \right)\delta
p_0 -(\rho+P)h'_0\nonumber\\
&& -\frac{2}{3}a_1'(\rho+P)\left[c_0-\frac{4}{5}c_1
  \left( a_1-\frac{3}{7}a_3 \right) \right] \nonumber\\
&&-\frac{12}{7}a_3'(\rho+P)c_1 \left(
\frac{2}{15}a_1-\frac{8}{15}a_3+\frac{10}{33}a_5 \right)\nonumber\\
&&-\frac{10}{11}a_5'(\rho+P)c_1 \left( \frac{4}{21}a_3-\frac{20}{39}a_5
\right)~,
\end{eqnarray}
Eqns. (\ref{eq0mh}) can be rearranged in the form
\begin{eqnarray}
&&m'_0-4\pi r^2\frac{\rho'}{P'}\delta
p_0=\frac{1}{3}(a'_1)^2 e^{-\lambda} +\frac{6}{7}(a'_3)^2
e^{-\lambda}+\frac{15}{11}(a'_5)^2 e^{-\lambda}\nonumber\\
&&+\frac{2}{3r^2}a_1^2+\frac{72}{7r^2}a_3^2
+\frac{450}{11r^2}a_5^2\nonumber\\
&&+\frac{e^{-\nu}}{4}\left[
\int_0^\pi \zeta_0^2 \left( |\psi/\bar{\psi}|-1 \right)^2
\Theta(|\psi/\bar{\psi}|-1) \frac{\psi^2}{\sin{\theta}} d\theta\right]
\;\; , \nonumber\\ 
&&\delta p'_0+\left[\frac{\nu'}{2}\left( \frac{\rho'}{P'}+1 \right)+4\pi r
e^\lambda (\rho+P)\right]\delta p_0 \nonumber\\
&&+e^{2\lambda}m_0(\rho+P)\left(
\frac{1}{r^2}+8\pi P\right)= \nonumber\\ &&
(\rho+P)\Bigg(-\frac{2}{3}a_1'\left[c_0-\frac{4}{5}c_1 \left( a_1-
\frac{3}{7}a_3\right) \right] \nonumber\\
&&-\frac{12}{7}a_3'c_1 \left(
\frac{2}{15}a_1-\frac{8}{15}a_3+\frac{10}{33}a_5 \right) \nonumber\\
&&-\frac{10}{11}a_5'c_1 \left(
\frac{4}{21}a_3-\frac{20}{39}a_5 \right)
-\frac{1}{3r}(a'_1)^2-\frac{6}{7r}(a'_3)^2 \nonumber\\
&&-\frac{15}{11r}(a'_5)^2+\frac{2e^{\lambda}}{3r^3}a_1^2 
+\frac{72e^{\lambda}}{7r^3}a_3^2+\frac{450e^{\lambda}}{11r^3}a_5^2
\nonumber\\
&&-\frac{e^{\lambda-\nu}}{4r}\left[ \int_0^\pi \zeta_0^2 \left(
|\psi/\bar{\psi}|-1 \right)^2 \Theta(|\psi/\bar{\psi}|-1)
\frac{\psi^2}{\sin{\theta}} d\theta \right] \Bigg) \;\; .\nonumber\\
\label{eqmp}
\end{eqnarray}
By imposing a regular behaviour at $r\simeq0$ we find
\begin{equation}
m_0(r\rightarrow 0)=Ar^3 \quad , \quad \delta p_0(r\rightarrow 0)=Cr^2
  \;\; ,
\end{equation}
where
\begin{eqnarray}
C&=&-\frac{2(P_c+\rho_c)\cdot (\alpha_1^2+\alpha_1 c_0)}{3+4\pi\left(
 r^2\frac{d\rho}{dP}\right)_c P_c} \nonumber\\
A&=&\frac{1}{3}\left[
 2\alpha_1^2+4\pi C\left( r^2\frac{d\rho}{dP}\right)_c \, \right] \;\; .
\end{eqnarray}
The subscript $c$ means that the quantity is evaluated at
$r\rightarrow0$. We remark that the solution of (\ref{eqmp}) does not
depend on new arbitrary constants. Outside the star, the equation for
$m_0$ reduces to
\begin{eqnarray}
m'_0&=&\left(
\frac{1}{3}(a_1')^2+\frac{6}{7}(a_3')^2+\frac{15}{11}(a_5')^2 \right)
\cdot\left( 1-\frac{2M}{r} \right)\nonumber\\
&&+\frac{2}{3r^2}a_1^2+\frac{72}{7r^2}a_3^2+\frac{450}{11r^2}a_5^2 \;\; .
\label{eqmext}
\end{eqnarray}
Solving 
(\ref{eqmp}), (\ref{eqmext}) we find $\delta M$ from (\ref{dMm}).
 
\end{document}